\documentstyle[preprint,prl,aps]{revtex}

\begin{document}
\thispagestyle{empty}


\footnotesep = 10pt
\addtolength{\skip\footins}{-3mm}

\preprint{
\vbox{\vskip -1.2cm \rightline{\tenrm $^{\dag}$SLAC-PUB-7391\phantom{.}}\vskip -.1cm 
\rightline{December 1996}\vskip -.1cm}}

\title{
\vskip -0.2cm
\baselineskip 14pt 
Measurements of the $\Delta$(1232) Transition Form Factor and the Ratio $\sigma_n/\sigma_p$ From Inelastic Electron-Proton and Electron-Deuteron Scattering$^{*}$ \vskip -0.3cm}

\footnotetext{\tenrm $^{*}$Work supported in part by
Department of Energy contract DE--AC03--76SF00515.}
\footnotetext{\tenrm $^{\dag}$This preprint supercedes SLAC-PUB-6305}

\author{\vskip -1cm
\baselineskip 12pt 
L.~M.~Stuart,$^{(2,4,10)}$
P.~E.~Bosted,$^{(1)}$                             
L.~Andivahis,$^{(1,a)}$ 
A.~Lung,$^{(1,b)}$                                       
J.~Alster,$^{(12)}$ 
R.~G.~Arnold,$^{(1)}$                                   
C.~C.~Chang,$^{(5)}$ 
F.~S.~Dietrich,$^{(4)}$
W.~R.~Dodge,$^{(7,c)}$
R.~Gearhart,$^{(10)}$ 
J.~Gomez,$^{(3)}$ 
K.~A.~Griffioen,$^{(8,d)}$ 
R.~S.~Hicks,$^{(6)}$ 
C.~E.~Hyde-Wright,$^{(13,e)}$
C.~Keppel,$^{(1,f)}$ 
S.~E.~Kuhn,$^{(11,e)}$ 
J.~Lichtenstadt,$^{(12)}$ 
R.~A.~Miskimen,$^{(6)}$ 
G.~A.~Peterson,$^{(6)}$
G.~G.~Petratos,$^{(9,g)}$ 
S.~E.~Rock,$^{(1)}$ 
S.~H.~Rokni,$^{(6,h)}$ 
W.~K.~Sakumoto,$^{(9)}$
M.~Spengos,$^{(1,i)}$ 
K.~Swartz,$^{(13)}$ 
Z.~Szalata,$^{(1)}$ 
L.~H.~Tao$^{(1)}$}

\address{\vskip -0.2cm
\baselineskip 12pt 
{~}\break
{$^{(1)}$The American University, Washington, D.C. 20016} \break  
{$^{(2)}$University of California, Davis, California 95616} \break                                                              
{$^{(3)}$CEBAF, Newport News, Virginia 23606} \break                               
{$^{(4)}$Lawrence Livermore National Laboratory, Livermore, California 94550} \break                                                   
{$^{(5)}$University of Maryland, College Park, Maryland 20742} \break                                                                
{$^{(6)}$University of Massachusetts, Amherst, Massachusetts 01003}  \break                                                                        
{$^{(7)}$National Institute of Standards and Technology, Gaithersburg, Maryland 20899} \break                                      
{$^{(8)}$University of Pennsylvania, Philadelphia, Pennsylvania 19104} \break                                                            
{$^{(9)}$University of Rochester, Rochester, New York 14627} \break                                                     
{$^{(10)}$Stanford Linear Accelerator Center, Stanford, California 94309 }  \break                              
{$^{(11)}$Stanford University, Stanford, California 94305} \break                                                               
{$^{(12)}$University of Tel-Aviv, Ramat Aviv, Tel-Aviv 69978, Israel} \break                                  
{$^{(13)}$University of Washington, Seattle, Washington 98195} \break                                         
\vskip -0.02cm}
\maketitle

\begin{minipage}{5.8in}  
\baselineskip 14pt 
Measurements of inclusive electron-scattering cross sections using hydrogen and deuterium targets in the region of the $\Delta$(1232) resonance are reported. A global fit to these new data and previous data in the resonance region is also reported for the proton. Transition form factors have been extracted from the proton cross sections for this experiment over the four-momentum transfer squared range $1.64 < Q^2 < 6.75$ (GeV/c)$^2$ and from previous data over the range $2.41 < Q^2 < 9.82$ (GeV/c)$^2$. The results confirm previous reports that the $\Delta$(1232) transition form factor decreases more rapidly with $Q^2$ than expected from perturbative QCD. The ratio of $\sigma _n /\sigma_p$ in the $\Delta$(1232) resonance   
region has been extracted from the deuteron data for this experiment in the range $1.64 < Q^2 < 3.75$ (GeV/c)$^2$ and for a previous experiment in the range $2.4 < Q^2 < 7.9$ (GeV/c)$^2$. A study has been made of the model dependence of these results. This ratio $\sigma _n/\sigma_p$ for $\Delta$(1232) production is slightly less than unity, while $\sigma _n/ \sigma_p$ for               
the nonresonant cross sections is approximately 0.5, which is consistent with deep inelastic scattering results.                            
\end{minipage}

\vskip 0.3cm \hskip 2.0in {\tenit  Submitted to Physical Review D}

\baselineskip = 24pt
\vfill\eject
\setcounter{page}{2}

\section{INTRODUCTION}
Understanding the structure of the nucleons and their excitations in terms of elementary constituents has been of fundamental interest for many years. In the limit of large four-momentum transfer squared $Q^2$, leading-order perturbative QCD (pQCD) is expected to be valid, but it is not clear how quickly in $Q^2$ the non-leading-order processes die off. The analysis of Stoler \cite{stoler} indicates that the $Q^{-4}$ form factor dependence expected from leading-order pQCD behavior may become evident as early as a few (GeV/c)$^2$ for the proton elastic form factor and for the transition form factors of the resonances at higher masses than the $\Delta$(1232). However, the transition form factor for the $\Delta$(1232) resonance does not display the expected leading-order pQCD behavior, even at $Q^2$ as high as 10 (GeV/c)$^2$. Instead, this form factor falls off more rapidly with $Q^2$ than expected. This implies that non-leading order processes are dominating the $\Delta$(1232) region while nearby regions exhibit leading-order pQCD behavior at the same $Q^2$. This anomaly makes the $\Delta$(1232) resonance an interesting candidate for further study. For both protons and neutrons there is a need for data on baryon excitation cross sections and transition form factors, especially at large $Q^2$, in order to provide for a better understanding of this effect and also to test alternate models.

In this experiment, NE11, performed at the Stanford Linear Accelerator Center (SLAC), measurements have been made of inclusive electron-scattering cross sections using hydrogen and deuterium targets in the region of the $\Delta$(1232) resonance. Also measured were the elastic form factors of the proton and the neutron \cite{bosta,lisa,lung} as well as the inelastic structure function, $\nu W_2$, and R = $\sigma_L /\sigma _T$ for electron-aluminum scattering \cite{bostb}. An overview of the experiment and the cross section results are given in Sections II and III. Section IV covers the development of a global fit for proton resonance cross sections using these new data along with some previous data. Also presented are transition form factors for the $\Delta$(1232) resonance extracted from the proton cross sections for the four-momentum transfer range $1.64 < Q^2 < 6.75$ (GeV/c)$^2$. In section V, results are shown for the ratio of $\sigma _n /\sigma_p$ extracted from the deuteron data for $1.64 < Q^2 < 3.75$ (GeV/c)$^2$ using several different Fermi smearing models and input assumptions to study the model dependence of the extraction. Results are also presented for the ratio of $\sigma _n /\sigma_p$ extracted from deuteron data taken during SLAC experiment E133 \cite{rock} for $2.42 < Q^2 < 7.86$ (GeV/c)$^2$.                                                                 

\section{EXPERIMENTAL SUMMARY}
  \subsection{Beam}
The electron beam, provided by the Nuclear Physics Injector at SLAC, operated at a beam pulse repetition rate of 120 Hz. A 5.5 GeV beam was produced with an average current of 5 $\mu$A and a pulse width of 2 $\mu$s. A 9.8 GeV beam was produced using the SLAC Energy Doubler with an average current of 1 $\mu$A and a beam pulse width of 0.15 $\mu$sec. The incident beam charge was measured by two independent toroidal charge monitors which were frequently calibrated by passing a known charge, generated using a precision capacitor, through a winding encircling the toroid. The two toroid measurements agreed to within $\pm$0.15\% and the absolute charge was measured to 0.5\%. The energy $E$ of the beam was monitored by a rotating flip-coil located within a dipole bending magnet identical to and in series with the dipole magnets used to steer the beam into the experimental area. Based on a calibration that required the elastic $ep$ cross-section peaks to be centered at a missing mass equal to the proton mass, the uncertainty in the beam energy was estimated to be 0.05\%. 

  \subsection{Targets}
The target assembly contained 15-cm-long liquid hydrogen and deuterium cells having 0.1-mm-thick aluminum endcaps and side walls. The endcap contribution to the measured cross sections was determined using a 1.8 mm aluminum target. In order to keep local density fluctuations below the level of 1\%, the liquid within the cells was circulated at a rate of 2 m/s. Target densities were determined using averaged temperature and pressure measurements from platinum resistors and vapor pressure bulbs located within the cells. These independent density measurements agreed to within 0.2\%. An absolute uncertainty was estimated at 0.9\% from uncertainties in cryogenic and resistor calibration data.                                                              

  \subsection{Spectrometer}
Scattered electrons were detected in the SLAC 8 GeV/c spectrometer \cite{kirk} located at forward scattering angles ranging from $\theta=13$ to $27^{\circ}$, and operated at central momenta ranging from 2.8 to 7.7 GeV/c. Uncertainties in the spectrometer central angle and momentum were $0.005^{\circ}$ and 0.05\% respectively. The detector package was designed for high electron detection efficiency in the presence of large pion backgrounds. It consisted of a gas threshold {$\rm\check C$}erenkov ({$\rm\check C$}) counter filled with 0.6 atmospheres of nitrogen operating at an efficiency of 99.0\%, ten planes of multi-wire proportional counters for particle tracking with a combined tracking efficiency of 99.9\%, and a lead glass shower counter array with an efficiency of 99.4\% and a resolution of $\pm 8\%/{\sqrt {E'}}$, where $E'$ is the energy of the scattered electron. The lead glass array was segmented into a 3.23 radiation-length pre-radiator (PR) followed by a total absorption (TA) counter composed of three layers, 6.8 radiation lengths each. For $E' <$ 4 GeV only two of the three layers were used. The detector package also used two layers of scintillators for triggering purposes (SF and SM). Resolutions for detected electron momentum and scattering angle were $\pm$0.15\% and $\pm$0.5 mr. \vskip .1in                                            
  \subsection{Electronics and Data Acquisition}
The data acquisition system used standard CAMAC and NIM electronic modules to process detector signals and to form event triggers. The trigger rate was restricted to one event per beam pulse due to limitations in the data logging rate. Additional triggers occurring in a beam pulse were counted in scalers for subsequent correction of the data. The event trigger required a beam gate and either an electron, pion, or a random trigger. The electron trigger consisted of either a three out of four coincidence between {$\rm\check C$}, PR, TA and SM or a two out of three coincidence between PR, SF and SM with {{$\rm\check C$}} always required. This trigger was designed for good efficiency over a large range in scattered electron momenta. The pion trigger required a coincidence between SF and SM, and was pre-scaled so that only a sampling of the pion background was analyzed.           
\section{DATA ANALYSIS}
  \subsection{Spectrometer Acceptance}
A Monte Carlo simulation of the spectrometer properties was used to determine the spectrometer acceptance as a function of relative momentum $\delta$, relative horizontal scattering angle $\Delta\theta$, and vertical scattering angle $\phi$. The input data for the simulation came from a survey of the spectrometer apertures and a TRANSPORT \cite{transport} calculation that agreed with floating-wire \cite{wirefloat} measurements of the spectrometer optical coefficients. In addition, corrections were calculated for momentum-dependent multiple scattering effects and changes in effective target length due to spectrometer rotation about the pivot.  
                                                                                
The $\delta$-dependence of the acceptance function was checked by comparing measurements of deuteron inelastic cross-section spectra taken at identical kinematics, except for the central spectrometer momenta, which differed by a few percent. The resulting smooth overlap between the spectra indicated that the $\delta$-dependence was understood. Elastic $ep$ cross sections were studied to verify that the acceptance function had the correct angular-dependence, namely, that there was no $\phi$-dependence and that the $\Delta\theta$ dependence did not differ from that predicted by a global fit of elastic cross sections covering a wide range of $\theta$.                       

  \subsection{Corrections to Data}
The measured counts were corrected for electronics and computer dead time and for detector inefficiencies. For each incident beam energy and scattering angle setting, cross sections were determined as a function of $E'$ by dividing the corrected counts by the number of incident electrons, the number of target nuclei per cm$^2$, and the spectrometer acceptance function. Corrections were also made for the small $\Delta\theta$ dependence of the cross section within the angular acceptance of the spectrometer. The average corrections for scattering from the aluminum endcaps amounted to 6\% for H$_2$ and 3\% for D$_2$. Pion contamination and $e^+/e^-$ pair production events at the target were found to be negligible for the kinematics of the data presented here.                       
                                                                                
Radiative corrections were also applied to the cross-section data. For the proton cross sections, the radiative tail for elastic scattering was calculated and subtracted using the formula of Tsai \cite{tsaia}, which is exact to lowest order in the fine structure constant $\alpha$. Multiplicative radiative corrections were then applied to both the remaining inelastic proton cross sections and the deuteron cross sections. These corrections were found using the peaking approximation formulas of Mo and Tsai \cite{tsaia,mo,tsaib} with additional corrections for vacuum polarization contributions from muon and quark loops \cite{bardin}. The final radiative corrections were calculated iteratively using a cross section model determined from a global fit as discussed below.

  \subsection{Inelastic Proton Cross Section Results}
Table I gives a listing of the final proton inelastic cross sections as a function of the kinematic variables $E',\ Q^2,$ and the longitudinal virtual photon polarization $\epsilon = [1 + 2(1 + \tau )$tan$^2(\theta/2)]^{-1}$, where $\tau$ is given by $(E - E')^2/Q^2$. Also shown are the multiplicative radiative corrections applied to the raw cross sections after elastic radiative tail subtraction, as well as the subtracted elastic radiative tail itself. The cross-section errors include statistical and point-to-point systematic uncertainties added in quadrature. The statistical uncertainties, typically $>2.5\%$, dominated. Point-to-point systematic errors were about 1\% from the combined uncertainties in beam energy (0.05\%), scattering angle (0.005$^\circ$), incident charge (0.15\%), detector efficiencies and electronic deadtime (0.25\%), radiative corrections (0.5\%), acceptance (0.3\%), and aluminum background subtraction (0.1\%). The overall normalization error was determined to be 1.8\% due to the combined absolute uncertainties in the incident charge (0.5\%), target density (0.9\%), target length (0.2\%), radiative corrections (1\%), and acceptance (1\%).                                
\section{PROTON CROSS SECTION FITS}
A global $Q^2$-dependent fit to the data was done in order to develop a good model of the resonance cross sections over a large kinematic range. To obtain the kinematic coverage in both $Q^2$ and $W^2$, additional data were included along with the data from this experiment. This fit was phenomenologically separated into resonant and nonresonant components which does not allow for the possibility of interference effects between the resonant and nonresonant processes, since this cannot be determined from inclusive experiments. Recent work \cite{nimai} investigating Bloom-Gilman duality \cite{bloom} in the $\Delta(1232)$ resonance region suggests that there may be common dynamics between the $\Delta$(1232) resonance and the underlying nonresonant background, possibly indicating that interference effects could be present, but more investigation is needed. The global fit was then used as an input for radiative correction calculations, for extracting information on the  $\Delta(1232)$ transition form factor as a function of $Q^2$, and for Fermi smearing calculations used in the $\sigma_n/\sigma_p$ extraction discussed below.          

  \subsection{Other Data Included in Analysis}
    \subsubsection{E133 Data}
In SLAC experiment E133 \cite{rock}, e-p and e-d cross sections for the resonance region were measured at a fixed scattering angle of $10^{\circ}$. The e-p data were in the range $2.4 < Q^2 < 9.8$ (GeV/c)$^2$, and the e-d data were in the range $2.4 < Q^2 < 7.9$ (GeV/c)$^2$. An error was found previously \cite{bostc} with the E133 data which affected the deduced beam energies. Accordingly, the E133 beam energies have been changed and the momenta have been slightly adjusted  within errors to align the quasielastic and $\Delta$(1232) resonance peaks at their appropriate masses. Also, the E133 data were normalized to the NE11 results which have smaller systematic errors. The normalization was found separately for the proton and deuteron data by minimizing the $\chi^2$ per degree of freedom (dof) for a global fit to the data over all Q$^2$. According to this procedure both the proton and deuteron normalization factors are 1.04. Note that the NE11 and E133 measurements were all made at forward angles.              

    \subsubsection{SLAC Deep Inelastic Data}
Global fits \cite{whitlowc,whitlowb,amaudruz,arneodo} have been made to deep inelastic cross-section data \cite{whitlowa,amaudruz,arneodo} for e-p and e-d scattering, using cross-normalized data sets with improved radiative corrections. The fits resulted in parameterizations of $R = \sigma_L/\sigma_T$ \cite{whitlowc} and the structure function $F_2(x,Q^2)$ \cite{whitlowb,amaudruz,arneodo} valid for missing mass squared $W^2 > 3.0$ GeV$^2$. It is naturally desirable that any global fit to the resonance region smoothly match fits to the deep inelastic region. To achieve this we used the same SLAC e-p data \cite{whitlowa} that were used in the deep inelastic analyses, subject to the restrictions $W^2 < 4.3$ GeV$^2$ and $Q^2 < 9.5$ (GeV/c)$^2$.

    \subsubsection{Resonance Region Data at low $Q^2$}
All inclusive e-p resonance region data measured up until the mid 1970's were evaluated and parameterized by Brasse {\it et al.} \cite{brasse}. This parameterization was used to generate cross-section spectra at five low-$Q^2$ values of 0.25, 0.50, 0.75, 1.0, and 1.3 GeV/c. These spectra were treated in the subsequent analysis as ``data''. The errors on these generated cross sections were assigned an additional normalization uncertainty of 7\% based on differences seen between these cross sections and recently reanalyzed SLAC resonance cross sections \cite{thia}. The $Q^2$ values of the generated spectra were chosen to be representative of the range of data originally parameterized.      

  \subsection{Cross Section Global Fit}
The differential inelastic proton cross section can be written as a sum of transverse and longitudinal terms:
\begin{equation}
{{d^2\sigma} \over {d\Omega dE'}}(E,E',\theta) = {{\alpha KE'} \over {4 \pi^2 Q^2E}} \ {2 \over {1-\epsilon}} [\sigma_T \ (W^2,Q^2) + \epsilon\sigma_L (W^2,Q^2)],
\end{equation}
where $K = (W^2 - M_p^2)/(2M_p)$ is the equivalent energy of a real photon needed to produce the same final mass state, and $M_p$ is the proton mass. These virtual photoabsorption cross sections can be expressed as:  
\begin{eqnarray}
\sigma_L &=&\sigma_L^{nr} = R\sigma_T^{nr},\nonumber\\ \sigma_T  
         &=& \sigma_T^{nr} + \sigma_{T\Delta} + \sigma_{T2} +\sigma_{T3}, 
\end{eqnarray}
where $\sigma_L^{nr}$ and $\sigma_T^{nr}$ are the longitudinal and transverse nonresonant background components, and $\sigma_{T\Delta}$, $\sigma_{T2}$, and $\sigma_{T3}$ are the transverse resonant components for the three dominant resonance regions, respectively. It was assumed that the longitudinal cross section for resonance production is zero, as indicated by the limited amount of available experimental data \cite{thia}, and that $R=\sigma_L^{nr}/\sigma_T^{nr}$ for the nonresonant cross sections could be parameterized by the expression $R = 0.25/{\sqrt{Q^2}}$ for $Q^2$ in (GeV/c)$^2$. This simple expression agrees reasonably well with a deep inelastic fit \cite{whitlowc} extrapolated to the kinematics of these data.                         

The quantity $\sigma_T^{nr}$ was described using a product of polynomials \cite{bartel} of the form
\begin{equation}                                  
{\sigma_T^{nr} \over G_D^2(Q^2)}= \sum^3_{i=1} C_i(Q^2)(W - W_{th})^{i-1/2},
\end{equation}                                                                    
where $W_{th}$ = $M_p + M_{\pi}$ is the pion production threshold given by the sum of the proton and pion masses, $C_i(Q^2)  = \sum_{n = 0}^{4}{Q^{(2*n)} C_{in}}$ are fitted $Q^2$-dependent amplitudes, and $G_D^2(Q^2) = 1/(1 + Q^2/0.71)^4$ with $Q^2$ in (GeV/c)$^2$ is the dipole form factor squared. The minimum number of fit parameters was chosen such that adequate fits to the data could be obtained over the desired kinematic range. 

The helicity-conserving and helicity-flip amplitudes for resonance production, $A_{1/2}(Q^2)$ and $A_{3/2}(Q^2)$ contribute only to the transverse virtual photoabsorption cross section. They can be combined to form the total transverse helicity amplitude   
\begin{equation}
\vert A_H(Q^2)\vert ^2 = \vert A_{1/2}(Q^2)\vert ^2 + \vert A_{3/2}(Q^2)\vert ^2. 
\end{equation}
Both the transition form factor and the transverse cross section for $\Delta(1232)$ production \cite{stoler,carlsonb} can be defined in terms of the helicity amplitudes:  
\begin{equation}
\vert F_\Delta (Q^2) \vert^2 = {1 \over 4 \pi\alpha}{2M_p \over Q^2} (M_\Delta ^2 - M_p^2) \vert A_H (Q^2) \vert^2,
\end{equation}                                                                   
\begin{equation}
\sigma_{T\Delta} ={2WM_p \over \Gamma_\Delta} \bigg({K_\Delta K_\Delta ^* \over K K^*}\bigg) {\Gamma _\gamma \Gamma _\pi \over (W^2 - M_\Delta ^2)^2 + (M_\Delta \Gamma _\pi)^2} \vert A_H (Q^2) \vert^2, 
\end{equation}                                           
where a relativistic Breit-Wigner \cite{stoler,walker} form has been used. The partial widths for the resonance are defined as 
\begin{equation}
\Gamma_\pi = \Gamma_\Delta \bigg[ {P_\pi^* \over P_{\pi\Delta}^*}\bigg]^3 \bigg[ {P_{\pi\Delta}^{*2} + X^2 \over P_\pi ^{*2}+ X^2} \bigg],\ \ \Gamma_\gamma = \Gamma_\Delta \bigg[                                 
{K^* \over K_{\Delta}^*}\bigg]^2 \bigg[ {K_{\Delta} ^{*2} + X^2 \over K^{*2} + X^2 }\bigg], \end{equation}  
where $M_{\Delta}$ and $\Gamma_\Delta$ are the $\Delta$(1232) resonance mass and width, K and K$^*$ are the equivalent energies of a real photon in the laboratory and center-of-mass frames needed to produce the final mass state W; $P_\pi^*$ is the decay pion momentum in the center-of-mass system; and a subscript of $\Delta$ on any of these quantities means that it is evaluated at the $\Delta$(1232) resonance peak. The parameter $X$ gives the mass variation of the resonance width. Photoproduction data fits yield a value $X= 0.16$ GeV for the $\Delta$(1232) resonance \cite{stein}. Results presented here are fairly insensitive to this parameter, but a ${\chi}^2$ best-fit to all the data yielded a value of $X=0.18$ GeV, which was used for all subsequent fitting. For the global fit, $F_{\Delta}(Q^2)$ was represented by a  $Q^2$-dependent fitting function given by ${\vert F_\Delta \vert^2 / G_D^2(Q^2)}= \sum_{n = 0}^{2}{Q^{(2*n)} \vert F_\Delta \vert_n^2}$.

There are many resonances which contribute to the resonance region beyond the $\Delta(1232)$ resonance, but the primary contributions can be separated into two mass regions which are denoted here as resonances 2 and 3. Resonance 2 is dominated \cite{brasseb} by the $D_{13}(1520)$ at low $Q^2$ and by the $S_{11}(1535)$ at high $Q^2$. These two resonances are of comparable magnitude around $Q^2= 1.3$ (GeV/c)$^2$. Resonance 3 is known to be dominated by the $F_{15}$(1680) for $Q^2<3.0$ (GeV/c)$^2$. For our purposes, it was adequate to represent these resonances by a simple nonrelativistic Breit-Wigner shape:     
\begin{equation}
{\sigma_{Ti} \over G_D^2(Q^2)} = A_i(Q^2){\Gamma_i \over (W - M_i)^2 +{1 \over 4}\Gamma_i ^2}. \end{equation}                                           
The index, i = 2 or 3 denotes the resonance 2 or 3, $A_i(Q^2) = \sum_{n = 0}^{1}{Q^{2n} A_{in}}$ are polynomial fits in $Q^2$, and $\Gamma_i$ and $M_i$ are the widths and masses of the resonances. Table II shows the resonance mass and width values used for fitting. These quantities were adjusted within reasonable limits to minimize the ${\chi}^2$ agreement between the global fit and the data. The best mass for resonance 3 was found to vary with $Q^2$, indicating that perhaps different resonances are contributing in this region at high $Q^2$. 
                    
Results for the resonance region global fit \cite{linda} are given in Table III and are shown in Fig. 1. These results are expected to be valid over the $Q^2$ range 0.3-10 (GeV/c)$^2$ and over the $W^2$ range from pion threshold out to 4.3 (GeV)$^2$. The fitted data were in units of $\mu$b, and the ${\chi}^2$ per 523 degrees of freedom was 1.56. Figure 1 shows three sample spectra containing both resonance and deep inelastic data. The deep inelastic data were from nearby kinematics and were kinematically corrected to the indicated kinematics using the resonance region global fit which is shown as a solid line. Also shown are the deep inelastic global fits given by the dashed curve \cite{whitlowb} and the dotted curve \cite{amaudruz}. Both of these used the same parameterization for $R$ \cite{whitlowc}. The new resonance region global fit is used for radiative corrections, for parameterizing the nonresonant component for the $\Delta (1232)$ transition form factor extraction, and for the analysis of the inelastic deuteron resonance data. 

  \subsection{$\Delta (1232)$ Transition Form Factors}
    \subsubsection{Form Factor Models}
Carlson and Poor \cite{carlsonb} have developed a distribution amplitude for the $\Delta$(1232) resonance using QCD sum rule constraints on the moments of the resonance wave function. A distribution amplitude is the momentum-space wave function which has been integrated over the transverse momenta. The $\Delta$(1232) distribution amplitude was then coupled with various nucleon distribution amplitudes in order to predict the magnitude of the transition form factors in the asymptotic limit of large $Q^2$. The nucleon distribution amplitudes used by Carlson and Poor include those of Chernyak and Zhitnitsky (CZ) \cite{cz}, King and Sachrajda (KS) \cite{ks}, and Gari and Stefanis (GS) \cite{gs}. Implicit in these asymptotic predictions is the assumption that $A_{3/2}(Q^2)$ can be neglected and that the transverse helicity amplitude $A_{1/2}(Q^2)$ dominates. Perturbative QCD predicts that $A_{1/2}(Q^2)$ falls as $1/Q^3$ and $A_{3/2}(Q^2)$ falls as $1/Q^5$ \cite{carlsona}, but this has not been established experimentally.                                                                          
                                                                                
The diquark model developed in part to describe the elastic electromagnetic nucleon form factors \cite{krollb}, was subsequently extended to the $\Delta$(1232) transition form factors as well \cite{krolla}. In this model, which inherently takes into account nonperturbative effects due to strong two-quark correlations, nucleons are considered as the combination of a quark and a spatially extended diquark. It has been suggested \cite{krolla} that the non-leading-order processes contributing to the rapid fall-off of the form factor of the $\Delta$ resonance are well-described within the framework of the diquark model. A model \cite{krolla} has been developed and tuned to agree with the E133 \cite{rock} results previously fit by Stoler \cite{stoler}. It has been shown \cite{bosta,lung} that this same model does not describe recent nucleon form factor measurements very well. Also, this model predicts a non-negligible contribution to the longitudinal resonance cross section which was assumed to be zero in this analysis. Data are needed to determine whether longitudinal resonance cross sections could be significant at large $Q^2$, and the diquark model needs to be re-examined to see if better agreement with nucleon form factors can be attained, and whether this has any effect on resonance form factor predictions. 

The recent heterotic calculations from Stefanis and Bergmann \cite{stefberg} are so-named because they are somewhat of a unification of the nucleon distribution amplitude models of Chernyak, Ogloblin, and Zhitnisky (COZ) \cite{COZ} and Gari and Stefanis \cite{gs} and the $\Delta$ distribution amplitude models of Carlson and Poor \cite{carlsonb} and Farrar {\it et al.} \cite{farrar}. In combining the various models to form the heterotic distribution amplitudes, an attempt has been made to retain the most promising features of the original models. The result is a new model which agrees better with existing data. 

    \subsubsection{Results For Transition Form Factors}
In order to obtain $\Delta$(1232) form factors as a function of $Q^2$, a fit was made to each individual resonance spectrum using the global nonresonant fit to describe the nonresonant background. The Brasse \cite{brasse} and E133 \cite{rock} data sets were fit for three resonance components as described for the global fits, except the fit coefficients were constants instead of polynomials in $Q^2$. Because the data from this experiment do not extend high enough in $W^2$, the global fit was also used to describe the net tail in the region of the $\Delta$(1232) resonance from higher-mass resonances. Thus, the only free parameter for the NE11 form factor fits was for the $\Delta(1232)$ form factor magnitude. The fits to all the individual spectra had a $\chi^2$/dof ranging from 0.14 to 4.1. Figure 2 shows sample NE11 data at four kinematic points. The curves indicate the $\Delta(1232)$ resonance (dotted), higher-mass resonance (dot-dash), nonresonant (dashed), and total (solid) cross sections. 

Figure 3 and Table IV give results for the $\Delta (1232)$ transition form factors relative to $\mu_p G_D(Q^2)$ where $\mu_p = 2.79$ is the proton anomalous magnetic moment in units of nuclear magnetons. The systematic uncertainties include both point-to-point and absolute systematic uncertainties, as well as modeling uncertainties due to the resonant and nonresonant global fit inputs. These results confirm previous results \cite{stoler} that the $\Delta (1232)$ transition form factor falls off more like $Q^{-6}$ than the expected pQCD result of $Q^{-4}$. A low-$Q^2$ multipole analysis \cite{davidson} shows that $\Delta$(1232) production is primarily a spin-flip transition and that the $A_{3/2}(Q^2)$ helicity amplitude is dominant. In contrast, the perturbative QCD expectation at high $Q^2$ is that $A_{1/2}(Q^2)$ is dominant. The data shown here are consistent with the helicity amplitude $A_{3/2}$ dominating for these kinematics and that the pQCD regime has not been reached. It is curious, however, that the nucleon form factor and the transition form factors for resonance regions 2 and 3 all seem to have the expected pQCD $Q^2$ behavior for the same $Q^2$ range \cite{stoler}. Note that the results given here are somewhat model dependent \cite{stoler,thia}. The differences between this analysis and that of Stoler \cite{stoler} include our use of the global fit to the nonresonant component rather than fitting this component separately for each cross-section spectrum, and also improved estimates of the systematic errors. Our method should yield a better overall representation of the nonresonant component, although the effect on the extracted form factors was small except for the spectra from this experiment which do not include data past the $\Delta(1232)$ resonance.

Also shown in Fig. 3 are the diquark model fit by Kroll {\it et al.} \cite{krolla}, one of three predictions from Stefanis and Bergmann \cite{stefberg}, and the three asymptotic predictions of Carlson and Poor \cite{carlsonb}. The Kroll curve agrees well with the data, but since the model was tuned to agree with the previous analysis \cite{stoler}, this is not surprising. The heterotic curve shown was calculated \cite{stefberg} using the heterotic nucleon and $\Delta$(1232) distribution amplitudes, and included corrections to estimate low $Q^2$ confinement effects as well as perturbative $Q^2$ evolution corrections.  This curve agrees with the data at high $Q^2$, but does not have the right shape at lower $Q^2$. The points (denoted by $*$) shown in Fig. 3 were evaluated at $Q^2 = 12$ (GeV/c)$^2$. The KS and CZ predictions are too low to describe the data, and the GS prediction is too high. New data at $Q^2$ larger than the results shown here would be valuable for determining whether the fall-off in $Q^2$ continues or the asymptotic limit has been reached. 

\section{DEUTERON CROSS SECTION ANALYSIS}
It is interesting to also consider neutron cross sections and to study how they compare with proton cross sections. Due to the lack of a free neutron target with sufficient density, the deuteron makes a good substitute. By combining proton and deuteron data we can learn how the two types of nucleons differ in their internal structure. This is typically done by extracting $\sigma_n/\sigma_p$ ratios. The goal of this analysis is to obtain results for $\sigma_n/\sigma_p$ for the $\Delta (1232)$ resonant and nonresonant cross section components separately and to compare them with expected results.

The analysis of the deuteron data is complicated by the fact that the composite nucleons are bound and have Fermi motion which smears the cross sections. In order to analyze the data, the components of the inelastic proton model cross sections (see Eq. 2) were smeared using several different prescriptions, and these smeared components were then fit to the deuteron data along with the quasielastic and meson-exchange components. For fitting purposes, all cross-section models and data were converted to a reduced form:
\begin{equation}
\sigma_R =  {d^2\sigma \over d\Omega dE'}\cdot{ \epsilon (1 + \tau) \over G_D^2(Q^2){\sigma_{Mott}}}= R_T(Q^2,W^2) + \epsilon R_L(Q^2,W^2), 
\end{equation}                            
where $\sigma_{Mott}$ = $\alpha^2$cos$^2(\theta/2)/4E^2$sin$^4(\theta/2)$, and $R_T(Q^2,W^2)$ and $R_L(Q^2,W^2)$ are the transverse and longitudinal components of $\sigma_R$ and are related to $\sigma_T$ and $\sigma_L$ by a common kinematic factor, $f= 4\alpha \pi^2 G_D^2 /K = \sigma_T/R_T = \sigma_L/R_L$.

  \subsection{Quasielastic Model}
Quasielastic scattering from the deuteron was described according to the Plane Wave Impulse Approximation (PWIA) of McGee and Durand \cite{mcgee}:                                                              
\begin{equation}
{d^2\sigma \over d\Omega dE'}(E,E',\theta ) = {M^2_p \over 2q} 
{E \over E'} (\sigma_p(E,\theta) + \sigma_n(E,\theta)) \int^{k_{max}}_{k_{min}}                                  
{[u^2(k) + w^2(k)]dk \over \sqrt{k^2 + M^2_p}}. 
\end{equation} 
In this expression the S-state and D-state momentum space amplitudes of the deuteron wave function are denoted by $u$ and $w$ respectively, and $q$ is the magnitude of the virtual photon three-momentum. The nucleon elastic cross sections are denoted by $\sigma_p$ and $\sigma_n$, and $k_{min}$ and $k_{max}$ are the minimum and maximum allowed values of the longitudinal Fermi momentum carried by the struck nucleon relative to the photon direction, as determined from energy conservation. The deuteron wave function was parameterized using the Paris potential \cite{paris}. Wave function modeling errors were estimated to be small using the alternate energy-independent Bonn \cite{bonn} and Reid \cite{reid} parameterizations in place of the Paris potential. Form factors for the elastic cross-section calculation were nominally taken from NE11 results \cite{bosta,lung}. However, a multiplicative fit parameter for the quasielastic contribution was included for two reasons. Firstly, the NE11 neutron form factor data are for $Q^2 \le 4.0$ (GeV/c)$^2$, whereas the E133 data analyzed here extend out to $Q^2 \simeq 8.0$ (GeV/c)$^2$, where the neutron elastic form factors have not been experimentally determined. Secondly, as discussed below, the inclusion of meson-exchange effects has a non-negligible effect on the neutron form factor extraction \cite{stuart}. Since further study is warranted on this issue, a fit parameter was allowed.

  \subsection{Meson-Exchange Currents and Final-State Interactions}
For the kinematics of the data presented here, no theoretical calculations exist for meson-exchange current (MEC) contributions to the cross sections. In lieu of these calculations the MEC contribution was estimated using calculations by Laget \cite{laget} for SLAC experiment NE4 \cite{arnold}. These calculations, which are restricted to $Q^2 \le 1.75$ (GeV/c)$^2$, include both MEC contributions and final state interactions (FSI), the latter being of relatively small consequence. The difference in calculated cross sections with and without the MEC and FSI contributions was fit as a function of {$W^2$} using a third-degree polynomial fit. This fit is shown in Fig. 4, where $\sigma_{MEC}$ is in reduced form, and the cross sections were assumed to be purely transverse so that $R_L = 0$. The fit shown in Fig. 4 was used for the shape of the MEC + FSI cross sections while the magnitude was adjusted by fitting the data. Since the relative contribution from MEC decreases with $Q^2$, this effect was only included for $Q^2 < 3.8$ (GeV/c)$^2$.                    

  \subsection{Fermi Smearing Methods}
Existing smearing techniques for nucleon cross-sections \cite{sfs,sargstrik} rely upon an incoherent impulse approximation in which only one of the two nucleons participates in the interaction. The non-interacting spectator nucleon is on its mass shell and is unaffected by the interaction, while the interacting nucleon is initially off-mass-shell, but is brought back on to the mass shell with the absorption of the virtual photon. Smearing method SM-I gives the smearing formulae in terms of light-cone variables \cite{sfs,sargstrik}. Smearing model SM-II involves a slight modification of the first model to account for a possible nuclear dependence from effects other than Fermi smearing in the deuteron at large $x$, where $x$, the Bjorken variable, is defined to be $Q^2/(2M_p\nu)$, and $\nu = E-E'$. This correction was calculated using a quark color screening model to explain nuclear dependencies in data. Smearing model SM-III is basically the Atwood-West \cite{atwood} smearing formalism, except that a correction has been made to the normalization of the deuteron wave function based on baryon charge conservation. Again, the Paris deuteron wave function parameterization was used. A study of results versus these smearing methods are presented below.                                           

  \subsection{Off-Shell Corrections to the Structure Functions}
The smeared structure functions are required to be off the mass shell. Since there are various prescriptions for relating on-shell and off-shell structure functions, the resultant systematic uncertainties were estimated using four different models. The simplest of these sets the off-shell structure functions equal to the on-shell structure functions OS-I. This assumption implies that the interacting nucleon is not far off the mass shell, presumably the case for the weakly bound deuteron system. Bodek {\it et al.,} \cite{bodek} require that the longitudinal virtual photoabsorption cross section $\sigma_{Ld}$ for the deuteron vanish in the limit $Q^2 \rightarrow 0$. This leads to two distinct off-shell corrections OS-II and OS-III (or any linear combination of the two corrections). There is some ambiguity in these corrections, and there is also no reason why the off-shell correction could not have a $Q^2$ dependence. Kusno and Moravczik \cite{kusno} assume that there is no off-shell correction applied to the transverse and longitudinal photoabsorption cross sections, $\sigma_T$ and $\sigma_L$. This assumption implies that there must be an off-shell correction OS-IV to the structure functions which are the quantities actually being smeared. This off-shell correction is also completely consistent with the constraint used for OS-II and OS-III that $\sigma_{Ld} \rightarrow 0$ as $Q^2 \rightarrow 0$, and there are no ambiguities. 

  \subsection{Cross-Section Fits}
The inelastic deuteron cross-section model was calculated using the global fit to the proton inelastic cross-section discussed previously, deuteron wave function models, Fermi smearing models, and off-shell corrections. For fitting purposes the resonant and nonresonant contributions to the structure functions were treated separately in the smearing formulae. The resonant component was separated into the $\Delta$(1232) contribution and a resonance tail contribution from the higher-mass resonance region 2. The data do not cover a large enough $W^2$ range to produce a good fit of the higher-mass resonance tail contributing at the $\Delta$(1232) resonance, so a fixed parameter $P$ was used. It is shown below that the results for $\sigma_n/\sigma_p$ are somewhat sensitive to the choice for this parameter. Also, it was assumed that the tail contribution from resonance region 3 could be neglected in the $\Delta$(1232) region.                                                            
                                                                                
The relationship between the structure functions used in the smearing formulae, $F_2$, and $W_2$, and the transverse photoabsorption cross section, is given by                                       
\begin{equation}
F_2(W^2,Q^2) = \nu W_2(W^2,Q^2) = {{\nu K(1 + R(W^2,Q^2))\sigma_T(W^2,Q^2)}  \over {4 \alpha \pi^2(1 + \tau)}}, 
\end{equation} 
where $R(W^2,Q^2) = \sigma_L/\sigma_T$ for the resonant and nonresonant components was defined earlier. The most important assumption made in the fitting procedure was that the shape of the smeared neutron cross sections in $E'$ for a fixed $E$ and $Q^2$ is the same as that of the smeared proton cross sections for each of the cross-section components.  The magnitudes of the neutron cross sections were allowed to differ from those of the proton cross sections. Thus, the deuteron cross-section data were fit using the smeared proton cross-section components as input. The coefficients found give information on the neutron contribution to the deuteron cross section, or equivalently information on the ratio of smeared cross sections, $\sigma_n/\sigma_p$, for each of the cross-section components. Three different fits were done. The first, Fit I, separately fits to each contribution as discussed in the text:
\begin{eqnarray}
\sigma_d &=& C_1\sigma_{quasi} + C_2\sigma_{_{MEC}} +   C_3\sigma_{p\_nres}^{sm} + C_4\sigma_{p\_\Delta}^{sm} +   P\sigma_{p\_2}^{sm}\nonumber\\
         &=& C_1\sigma_{quasi} + C_2\sigma_{_{MEC}} +   {1\over2}(\sigma_{p\_nres}^{sm} + \sigma_{n\_nres}^{sm} + \sigma_{p\_\Delta}^{sm} + \sigma_{n\_\Delta}^{sm}) +   P\sigma_{p\_2}^{sm}, 
\end{eqnarray}                                      
where $\sigma_d$ is the measured per nucleon deuteron cross section, $\sigma_{quasi}$ and $\sigma_{MEC}$ are the model quasielastic and MEC reduced cross sections for the deuteron, and $C_1$, $C_2$, $C_3$, and $C_4$ are fit parameters. The superscript $sm$ means ``smeared'', subscripts $p\_nres$ and $n\_nres$ refer to the proton and neutron nonresonant contributions, $p\_\Delta$ and $n\_\Delta$ refers to the proton and neutron $\Delta$(1232) resonant contributions, and $p\_2$ refers to the higher-mass resonance 2 contribution. It is straightforward to show from Eq. 12 with the assumptions given above that ${(\sigma_n/\sigma_p)}_{nres} = (2C_3 - 1)$, 
${(\sigma_n/\sigma_p)}_{\Delta} = (2C_4 - 1)$, and ${(\sigma_n/\sigma_p)}_{2} = (2P - 1)$, where ${(\sigma_n/\sigma_p)}_{2}$ represents the ${(\sigma_n/\sigma_p)}$ ratio for the higher-mass resonance 2.
The second fit, Fit II, is similar to Fit I except the ratio ${\sigma_n/\sigma_p}_{\Delta}$ was forced to be unity. The motivation for this is discussed below. Finally, Fit III was performed such that the smeared proton resonance and nonresonant components were not separately fit, but first combined to yield ${(\sigma_n/\sigma_p)}$ information for the total cross section in the region of the $\Delta$(1232) resonance:
\begin{equation}
\sigma_d = C'_1\sigma_{quasi} + C'_2\sigma_{_{MEC}} +   C'_3(\sigma_{p\_nres}^{sm} + \sigma_{p\_\Delta}^{sm}) +   P\sigma_{p\_2}^{sm},
\end{equation}                                      
where ${(\sigma_n/\sigma_p)}_{total} = (2C'_3 - 1)$.

  \subsection{Results for $\sigma_n/\sigma_p$}
The NE11 and E133 deuteron cross sections are shown in Figs. 5 and 6, respectively. Contributions from the cross-section components are shown as well as the total model cross section. Contributions from the second resonance region and MEC are not shown in Figs. 5 and 6 because they are too small to be seen clearly. These results are for the model choices defined by the Paris deuteron wave function, OS-IV for the off-shell correction, SM-I for the smearing model, and ${(\sigma_n/\sigma_p)}_{2}=0.3$. These conditions are hereafter referred to as ``standard'', and a study of the model dependence is shown below. 

Upper and lower limits derived on ${(\sigma_n/\sigma_p)}_{2}$ from SU(6) symmetry assumptions \cite{close} indicate that this ratio should be between zero and unity. Existing low-Q$^2$ data \cite{kobb} shown in Fig. 7 seem to indicate that the ratio is decreasing with increasing Q$^2$. A relativistic constituent quark model prediction \cite{warns} for the S$_{11}(1535)$ resonance is approximately level at ${(\sigma_n/\sigma_p)} = 0.3$ for $Q^2 > 1.5$ (GeV/c)$^2$, the kinematics of the present data. This model also agrees well with the low $Q^2$ data, although these data do contain contributions from both the D$_{13}(1520)$ and S$_{11}(1535)$ resonances. Nevertheless, the S$_{11}(1535)$ is the dominant resonance for $Q^2 > 1.3$ (GeV/c)$^2$ \cite{brasseb}. The sensitivity of the results to ${(\sigma_n/\sigma_p)}_{2}$ is examined below.

The $\chi^2$/dof for the ``standard'' individual fits ranged from 0.5 to 2.2 for Fit I. The $\chi^2$/dof were only slightly worse for fits II and III. Figure 8 shows the results for $\sigma_n/\sigma_p$ for the nonresonant and $\Delta$ (1232) excitation cross sections under ``standard'' conditions for Fit I, and Table V shows results for all three fits.  In Fig. 8, the smaller error bar is statistical, and the larger is the total uncertainty including experimental systematic, uncertainties due to the resonant and nonresonant global fit inputs, and modeling errors which are discussed more below.

A study was made of the model dependence of the results. Figures 9 and 10 show the results from Fit I for the $\Delta(1232)$ and the nonresonant background, respectively. This study was done for different deuteron wave functions, smearing models, off-shell corrections, and the assumed amount of higher-mass resonance contribution. The errors shown are the statistical uncertainties for the ``standard'' data points (square points) only. The dependence on off-shell correction and deuteron wave functions is small compared to the statistical uncertainties; the dependence on the smearing model is greater, but only exceeds the statistical uncertainty for the higher $Q^2$ E133 points; and the dependence on the assumed ${(\sigma_n/\sigma_p)}_{2}$ contribution is typically the largest uncertainty, especially at high $Q^2$. These modeling trends are similar for fits II and III.

It is usually assumed that the proton and neutron $\Delta$(1232) resonant amplitudes are entirely isovector transitions ($\Delta I = 1$). There can obviously be no isoscalar transitions from the I=1/2 nucleon ground state to the I=3/2 $\Delta$(1232) resonance. However, the electromagnetic current could contain an isotensor contribution which would allow $\Delta I = 2$ transitions as well. Thus, these resonant amplitudes can be decomposed into isovector and isotensor ($A_T$) components \cite{gittelman,bleckwenn}, such that the ratio $\sigma_n/\sigma_p$ can be expressed
\begin{equation}
{\sigma_n\over \sigma_p} = {(A_V + A_T)^2 \over (A_V - A_T)^2}.
\end{equation}
In the absence of an isotensor contribution, the ratio $\sigma_n/\sigma_p$ should be unity. Low $Q^2$ electroproduction data \cite{kobb,bleckwenn} indicates that the isotensor contribution is small, but there is a systematic trend for the $\sigma_n/\sigma_p$ data to be less than unity. An average over all $Q^2$ of the low $Q^2$ K\"obberling data \cite{kobb} yields $\sigma_n/\sigma_p=0.91\pm0.03$. The resonance results shown in Fig. 8a are consistent with the low $Q^2$ data and also show a trend to be less than unity. An average over all $Q^2$ of the NE11 and the E133 data yields $\sigma_n/\sigma_p=0.72\pm0.07$. It is also interesting to note that there is no observed $Q^2$ dependence to the $(\sigma_n/\sigma_p)_{\Delta}$ data. Although the errors are large, this implies that the neutron $\Delta$(1232) transition form factor exhibits a similar behavior in $Q^2$ to that of the proton transition form factor.

The results for $\sigma_n/\sigma_p$ for the nonresonant cross sections are expected to be consistent with deep-inelastic results where the resonant contributions have died away. Figure 8b shows two curves which give some indication of what the expected results should be. The upper curve is the ratio of $F_2^n(x,Q^2)/F_2^p(x,Q^2)$ evaluated at a fixed $W^2$ = 4 GeV$^2$ as  given by a fit to deep inelastic data \cite{amaudruz}. The second curve is the expected SU(6) limit of $F_2^n(x)/F_2^p(x)$ as $x$ $\rightarrow$ 1 \cite{carlitz}. The ratio for $\sigma_n/\sigma_p$ for the nonresonant background is likely to fall somewhere between the two curves. The data are a little high, but within errors are consistent with this expectation. For Fit II where ${(\sigma_n/\sigma_p)}_{\Delta}$ is forced to be unity, the results for ${(\sigma_n/\sigma_p)}_{nres}$ are generally lower than the Fit I results. The Fit II results are shown in Table V.

The total ratio ${(\sigma_n/\sigma_p)}_{total}$ was determined from Fit III, and is given in Table V. The NE11 and E133 results are consistent in the overlap region. Note that the E133 results for ${(\sigma_n/\sigma_p)}_{total}$ were previously published \cite{rock} from an independent analysis, and are consistent with the results presented here, although the new results have a larger estimate for the modeling uncertainty.

\section{SUMMARY AND CONCLUSIONS}
A global fit to proton inelastic cross sections, which phenomenologically separates the nonresonant and resonant components, provides a reliable model over the range $0.3 < Q^2 < 10$ (GeV/c)$^2$ and for W$^2$ between pion threshold and 4.3 (GeV)$^2$. This resonance global fit was designed to smoothly match the deep-inelastic SLAC global fit \cite{whitlowb} at W$^2 \approx 4$ (GeV)$^2$. Using the resonance global fit, new results have been extracted for the $\Delta(1232)$ transition form factors over the range $1.64 < Q^2 < 6.75$ (GeV/c)$^2$. These results confirm that the $\Delta (1232)$ transition form factor decreases with increasing Q$^2$ faster than that expected from pQCD \cite{stoler}.

New results have also been presented from Fit I where $\sigma_n/\sigma_p$ ratios are extracted separately for the $\Delta$(1232) resonance and the nonresonant background from inclusive electron-deuteron scattering cross sections in the resonance region. The results are consistent with $(\sigma_n/\sigma_p)_{\Delta}$ being slightly less than unity as previous data \cite{kobb} also seem to indicate, and there is no notable $Q^2$ dependence to this quantity. This implies, with large errors, that the neutron $\Delta$(1232) transition form factor has a similar $Q^2$ dependence to that of the proton. The results shown for $(\sigma_n/\sigma_p)_{nres} \sim 0.5$ are consistent with deep inelastic results. The model dependence of the $\sigma_n/\sigma_p$ ratio extraction has been studied, and is the dominant uncertainty for the high $Q^2$ data.

\eject                                                                                                    

\begin{tabular}{|c c p{0.6in} c c c|}
\hline
\multicolumn{6}{|p{4.3in}|}{Table I. Cross sections for inelastic electron scattering from protons for SLAC experiment NE11. The errors shown include statistical and point-to-point systematic uncertainties  added in quadrature, but do not include an overall normalization error of about 1.8\%. Also shown are RC, the applied multiplicative radiative corrections to the raw cross sections after elastic tail subtraction, and RT, the subtracted elastic radiative tail. } \\  \hline
 & & & & & \\
$E^{\,\prime}$ & $Q^2$ &  \ \ $\epsilon$\ \  & RC & RT & $d\sigma /d\Omega dE^{\,\prime}$ \\
 (GeV) & \ (GeV/c)$^2$\  & & & \ (nb/sr-GeV)\  & \ (nb/sr-GeV)\   \\ 
 & & & & & \\ \hline
 & & & & & \\ 
\multicolumn{6}{|c|}{$E=5.507$ GeV \ \ $\theta=15.146^\circ$} \\
 & & & & & \\ 
 4.437 & 1.698 & 0.944 &  1.82 &   7.58  & $  4.21\pm  1.04$ \\ 
 4.429 & 1.694 & 0.944 &  1.73 &   7.20  & $  4.85\pm  1.06$ \\ 
 4.420 & 1.691 & 0.943 &  1.68 &   6.86  & $  5.86\pm  1.08$ \\ 
 4.411 & 1.687 & 0.943 &  1.64 &   6.57  & $  8.76\pm  1.13$ \\ 
 4.402 & 1.684 & 0.943 &  1.62 &  6.3  & $12.1\pm 1.2{\phantom{0}} $ \\  
 4.393 & 1.681 & 0.942 &  1.60 &  6.0  & $15.4\pm 1.2{\phantom{0}} $ \\  
 4.384 & 1.677 & 0.942 &  1.59 &  5.8  & $17.5\pm 1.3{\phantom{0}} $ \\  
 4.375 & 1.674 & 0.941 &  1.58 &  5.6  & $22.5\pm 1.4{\phantom{0}} $ \\  
 4.367 & 1.670 & 0.941 &  1.57 &  5.4  & $26.4\pm 1.5{\phantom{0}} $ \\  
 4.358 & 1.667 & 0.940 &  1.56 &  5.3  & $32.9\pm 1.6{\phantom{0}} $ \\  
 4.349 & 1.664 & 0.940 &  1.56 &  5.1  & $37.6\pm 1.7{\phantom{0}} $ \\  
 4.340 & 1.660 & 0.940 &  1.55 &  5.0  & $48.5\pm 1.9{\phantom{0}} $ \\  
 4.331 & 1.657 & 0.939 &  1.53 &  4.8  & $56.1\pm 2.0{\phantom{0}} $ \\  
 4.322 & 1.654 & 0.939 &  1.51 &  4.7  & $62.2\pm 2.1{\phantom{0}} $ \\  
 4.313 & 1.650 & 0.938 &  1.49 &  4.6  & $64.3\pm 2.1{\phantom{0}} $ \\  
 4.304 & 1.647 & 0.938 &  1.45 &  4.5  & $68.1\pm 2.2{\phantom{0}} $ \\  
 4.296 & 1.643 & 0.937 &  1.42 &  4.4  & $65.7\pm 2.1{\phantom{0}} $ \\  
 4.287 & 1.640 & 0.937 &  1.38 &  4.3  & $63.6\pm 2.1{\phantom{0}} $ \\  
 4.278 & 1.637 & 0.936 &  1.35 &  4.2  & $58.4\pm 2.1{\phantom{0}} $ \\  
 4.269 & 1.633 & 0.936 &  1.31 &  4.1  & $57.1\pm 2.1{\phantom{0}} $ \\  
 4.260 & 1.630 & 0.935 &  1.29 &  4.0  & $54.6\pm 2.1{\phantom{0}} $ \\  
 4.252 & 1.626 & 0.935 &  1.26 &  4.0  & $50.3\pm 2.2{\phantom{0}} $ \\  
 4.243 & 1.623 & 0.934 &  1.24 &  3.9  & $48.0\pm 2.3{\phantom{0}} $ \\  
 4.234 & 1.620 & 0.934 &  1.23 &  3.8  & $52.1\pm 2.7{\phantom{0}} $ \\  
 4.225 & 1.616 & 0.933 &  1.22 &  3.8  & $48.1\pm 2.9{\phantom{0}} $ \\  
 4.216 & 1.613 & 0.933 &  1.21 &  3.7  & $45.8\pm 3.6{\phantom{0}} $ \\  
 4.207 & 1.610 & 0.932 &  1.20 &  3.7  & $46.4\pm 4.7{\phantom{0}} $ \\  
 4.198 & 1.606 & 0.932 &  1.19 &  3.6  & $50.8\pm 9.2{\phantom{0}} $ \\  \hline
\end{tabular}
\eject
\begin{tabular}{|c c p{0.6in} c c c|}
\hline
\multicolumn{6}{|p{4.3in}|}{Table I. Continued. } \\  \hline
 & & & & & \\
$E^{\,\prime}$ & $Q^2$ &  \ \ $\epsilon$\ \  & RC & RT & $d\sigma /d\Omega dE^{\,\prime}$ \\
 (GeV) & \ (GeV/c)$^2$\  & & & \ (nb/sr-GeV)\  & \ (nb/sr-GeV)\   \\ 
 & & & & & \\ \hline
 & & & & & \\ 
\multicolumn{6}{|c|}{$E=5.507$ GeV \ \ $\theta=18.981^\circ$} \\
 & & & & & \\ 
 4.060 & 2.431 & 0.906 &  2.10 &  1.450  & $ 0.549\pm 0.144$ \\ 
 4.048 & 2.424 & 0.905 &  1.80 &  1.337  & $ 0.814\pm 0.129$ \\ 
 4.035 & 2.417 & 0.904 &  1.70 &   1.24  & $  1.21\pm  0.13$ \\ 
 4.024 & 2.409 & 0.903 &  1.64 &   1.17  & $  1.76\pm  0.14$ \\ 
 4.011 & 2.402 & 0.903 &  1.61 &   1.10  & $  2.13\pm  0.15$ \\ 
 3.999 & 2.395 & 0.902 &  1.59 &   1.04  & $  2.94\pm  0.18$ \\ 
 3.987 & 2.388 & 0.901 &  1.58 &   0.99  & $  4.20\pm  0.20$ \\ 
 3.975 & 2.380 & 0.900 &  1.57 &   0.95  & $  6.19\pm  0.23$ \\ 
 3.963 & 2.373 & 0.899 &  1.55 &   0.91  & $  7.57\pm  0.25$ \\ 
 3.951 & 2.366 & 0.898 &  1.53 &   0.87  & $  9.30\pm  0.27$ \\ 
 3.938 & 2.359 & 0.898 &  1.50 &  0.8  & $10.8\pm 0.3{\phantom{0}} $ \\  
 3.927 & 2.351 & 0.897 &  1.45 &  0.8  & $11.4\pm 0.3{\phantom{0}} $ \\  
 3.914 & 2.344 & 0.896 &  1.40 &  0.8  & $11.3\pm 0.3{\phantom{0}} $ \\  
 3.902 & 2.337 & 0.895 &  1.35 &  0.8  & $10.7\pm 0.3{\phantom{0}} $ \\  
 3.890 & 2.330 & 0.894 &  1.31 &   0.75  & $  9.99\pm  0.30$ \\ 
 3.878 & 2.322 & 0.893 &  1.28 &   0.73  & $  9.38\pm  0.32$ \\ 
 3.866 & 2.315 & 0.892 &  1.26 &   0.71  & $  9.53\pm  0.37$ \\ 
 3.854 & 2.308 & 0.891 &  1.24 &   0.70  & $  9.08\pm  0.47$ \\ 
 3.842 & 2.300 & 0.890 &  1.23 &   0.68  & $  8.84\pm  0.64$ \\ 
 3.830 & 2.293 & 0.889 &  1.22 &   0.67  & $  7.57\pm  1.26$ \\ 
 & & & & & \\ 
\multicolumn{6}{|c|}{$E=5.507$ GeV \ \ $\theta=22.805^\circ$} \\
 & & & & & \\ 
 3.659 & 3.149 & 0.855 &  1.78 &  0.327  & $ 0.187\pm 0.026$ \\ 
 3.644 & 3.137 & 0.854 &  1.66 &  0.299  & $ 0.334\pm 0.027$ \\ 
 3.629 & 3.124 & 0.852 &  1.61 &  0.276  & $ 0.490\pm 0.033$ \\ 
 3.614 & 3.112 & 0.851 &  1.58 &  0.258  & $ 0.714\pm 0.038$ \\ 
 3.600 & 3.099 & 0.850 &  1.56 &   0.24  & $  1.19\pm  0.05$ \\ 
 3.585 & 3.087 & 0.848 &  1.55 &   0.23  & $  1.70\pm  0.06$ \\ 
 3.571 & 3.074 & 0.847 &  1.52 &   0.22  & $  2.28\pm  0.07$ \\ 
 3.556 & 3.061 & 0.846 &  1.47 &   0.21  & $  2.64\pm  0.07$ \\ 
 3.541 & 3.049 & 0.844 &  1.41 &   0.20  & $  2.68\pm  0.07$ \\ 
 3.527 & 3.036 & 0.843 &  1.35 &   0.19  & $  2.69\pm  0.07$ \\ 
 3.512 & 3.024 & 0.841 &  1.30 &   0.19  & $  2.54\pm  0.08$ \\ 
 3.498 & 3.011 & 0.840 &  1.27 &   0.18  & $  2.51\pm  0.09$ \\ \hline
\end{tabular}
\eject
\begin{tabular}{|c c p{0.6in} c c c|}
\hline
\multicolumn{6}{|p{4.3in}|}{Table I. Continued. } \\  \hline
 & & & & & \\
$E^{\,\prime}$ & $Q^2$ &  \ \ $\epsilon$\ \  & RC & RT & $d\sigma /d\Omega dE^{\,\prime}$ \\
 (GeV) & \ (GeV/c)$^2$\  & & & \ (nb/sr-GeV)\  & \ (nb/sr-GeV)\   \\ 
 & & & & & \\ \hline
 & & & & & \\ 
 3.483 & 2.999 & 0.839 &  1.26 &   0.18  & $  2.66\pm  0.12$ \\ 
 3.468 & 2.986 & 0.837 &  1.25 &   0.18  & $  2.26\pm  0.18$ \\ 
 3.454 & 2.973 & 0.836 &  1.24 &   0.17  & $  2.08\pm  0.57$ \\ 
 & & & & & \\ 
\multicolumn{6}{|c|}{$E=5.507$ GeV \ \ $\theta=26.823^\circ$} \\
 & & & & & \\ 
 3.279 & 3.886 & 0.794 &  1.97 & 0.1056  & $0.0376\pm0.0086$ \\ 
 3.263 & 3.867 & 0.793 &  1.69 & 0.0938  & $0.0820\pm0.0080$ \\ 
 3.247 & 3.847 & 0.791 &  1.61 &  0.085  & $ 0.130\pm 0.009$ \\ 
 3.230 & 3.828 & 0.789 &  1.57 &  0.078  & $ 0.231\pm 0.011$ \\ 
 3.214 & 3.809 & 0.787 &  1.55 &  0.073  & $ 0.382\pm 0.013$ \\ 
 3.198 & 3.789 & 0.785 &  1.52 &  0.069  & $ 0.557\pm 0.015$ \\ 
 3.181 & 3.770 & 0.783 &  1.48 &  0.065  & $ 0.744\pm 0.017$ \\ 
 3.165 & 3.750 & 0.781 &  1.40 &  0.063  & $ 0.790\pm 0.018$ \\ 
 3.149 & 3.731 & 0.779 &  1.33 &  0.060  & $ 0.808\pm 0.019$ \\ 
 3.132 & 3.712 & 0.777 &  1.29 &  0.058  & $ 0.802\pm 0.021$ \\ 
 3.116 & 3.692 & 0.775 &  1.27 &  0.057  & $ 0.818\pm 0.028$ \\ 
 3.100 & 3.673 & 0.773 &  1.26 &  0.055  & $ 0.851\pm 0.046$ \\ 
 3.083 & 3.654 & 0.771 &  1.26 &  0.054  & $ 0.575\pm 0.171$ \\ 
 & & & & & \\ 
\multicolumn{6}{|c|}{$E=9.800$ GeV \ \ $\theta=13.248^\circ$} \\
 & & & & & \\ 
 7.545 & 3.936 & 0.942 &  2.11 &  0.427  & $ 0.234\pm 0.062$ \\ 
 7.525 & 3.925 & 0.941 &  1.80 &  0.374  & $ 0.391\pm 0.055$ \\ 
 7.504 & 3.914 & 0.940 &  1.72 &  0.335  & $ 0.672\pm 0.060$ \\ 
 7.483 & 3.903 & 0.940 &  1.67 &   0.30  & $  1.07\pm  0.07$ \\ 
 7.462 & 3.892 & 0.939 &  1.65 &   0.28  & $  1.76\pm  0.08$ \\ 
 7.441 & 3.882 & 0.938 &  1.63 &   0.26  & $  2.64\pm  0.10$ \\ 
 7.420 & 3.871 & 0.938 &  1.58 &   0.24  & $  3.43\pm  0.12$ \\ 
 7.399 & 3.860 & 0.937 &  1.50 &   0.23  & $  3.53\pm  0.13$ \\ 
 7.379 & 3.849 & 0.936 &  1.43 &   0.22  & $  3.53\pm  0.15$ \\ 
 7.358 & 3.838 & 0.936 &  1.39 &   0.20  & $  3.29\pm  0.18$ \\ 
 7.337 & 3.827 & 0.935 &  1.36 &   0.20  & $  3.57\pm  0.24$ \\ 
 7.316 & 3.816 & 0.934 &  1.35 &   0.19  & $  3.34\pm  0.36$ \\ \hline 
\end{tabular}
\eject
\begin{tabular}{|c c p{0.6in} c c c|}
\hline
\multicolumn{6}{|p{4.3in}|}{Table I. Continued. } \\  \hline
 & & & & & \\
$E^{\,\prime}$ & $Q^2$ &  \ \ $\epsilon$\ \  & RC & RT & $d\sigma /d\Omega dE^{\,\prime}$ \\
 (GeV) & \ (GeV/c)$^2$\  & & & \ (nb/sr-GeV)\  & \ (nb/sr-GeV)\   \\ 
 & & & & & \\ \hline
 & & & & & \\ 
\multicolumn{6}{|c|}{$E=9.800$ GeV \ \ $\theta=15.367^\circ$} \\
 & & & & & \\ 
 7.016 & 4.916 & 0.914 &  1.99 & 0.1150  & $0.0844\pm0.0190$ \\ 
 6.988 & 4.896 & 0.913 &  1.74 &  0.096  & $ 0.183\pm 0.019$ \\ 
 6.958 & 4.876 & 0.912 &  1.67 &  0.083  & $ 0.360\pm 0.023$ \\ 
 6.929 & 4.855 & 0.911 &  1.63 &  0.074  & $ 0.595\pm 0.028$ \\ 
 6.900 & 4.835 & 0.909 &  1.56 &  0.067  & $ 0.880\pm 0.039$ \\ 
 6.870 & 4.815 & 0.908 &  1.46 &   0.06  & $  1.04\pm  0.05$ \\ 
 6.842 & 4.794 & 0.907 &  1.40 &  0.057  & $ 0.994\pm 0.062$ \\ 
 6.812 & 4.774 & 0.905 &  1.38 &  0.054  & $ 0.908\pm 0.100$ \\ 
 & & & & & \\ 
\multicolumn{6}{|c|}{$E=9.800$ GeV \ \ $\theta=17.516^\circ$} \\
 & & & & & \\ 
 6.492 & 5.900 & 0.881 &  2.00 & 0.0381  & $0.0195\pm0.0070$ \\ 
 6.466 & 5.875 & 0.879 &  1.73 & 0.0319  & $0.0755\pm0.0078$ \\ 
 6.438 & 5.851 & 0.878 &  1.66 &  0.028  & $ 0.115\pm 0.009$ \\ 
 6.412 & 5.826 & 0.876 &  1.62 &  0.025  & $ 0.203\pm 0.011$ \\ 
 6.384 & 5.802 & 0.875 &  1.56 &  0.022  & $ 0.284\pm 0.014$ \\ 
 6.358 & 5.777 & 0.873 &  1.47 &  0.021  & $ 0.317\pm 0.018$ \\ 
 6.330 & 5.753 & 0.872 &  1.42 &  0.019  & $ 0.342\pm 0.023$ \\ 
 6.304 & 5.728 & 0.871 &  1.40 &  0.018  & $ 0.315\pm 0.033$ \\ 
 6.277 & 5.704 & 0.869 &  1.39 &  0.017  & $ 0.370\pm 0.114$ \\ 
 & & & & & \\ 
\multicolumn{6}{|c|}{$E=9.800$ GeV \ \ $\theta=19.753^\circ$} \\
 & & & & & \\ 
 5.944 & 6.855 & 0.839 &  1.73 & 0.0120  & $0.0330\pm0.0055$ \\ 
 5.919 & 6.826 & 0.837 &  1.65 & 0.0105  & $0.0382\pm0.0058$ \\ 
 5.895 & 6.797 & 0.836 &  1.60 & 0.0093  & $0.0582\pm0.0073$ \\ 
 5.869 & 6.769 & 0.834 &  1.55 &  0.009  & $ 0.102\pm 0.010$ \\ 
 5.845 & 6.740 & 0.832 &  1.47 &  0.008  & $ 0.110\pm 0.011$ \\ 
 5.820 & 6.712 & 0.831 &  1.43 &  0.007  & $ 0.125\pm 0.015$ \\ 
 5.795 & 6.683 & 0.829 &  1.41 &  0.007  & $ 0.154\pm 0.025$ \\ 
 5.770 & 6.655 & 0.827 &  1.40 &  0.007  & $ 0.168\pm 0.052$ \\ \hline 
\end{tabular}
\eject

\begin{tabular}{|c p{0.8in} c|}
\hline
\multicolumn{3}{|p{3.7in}|}{Table II. Resonance widths and masses used in fits. The index i = $\Delta$, 2, or 3 denotes the $\Delta$(1232) resonance and resonances 2 and 3, respectively. $Q^2$ is in units of (GeV/c)$^2$ } \\  \hline
 & & \\
 $i$ &  \hfill $\Gamma_i$ (GeV) &$M_i$ (GeV)  \\
 & & \\ \hline
 & & \\ 
 $\Delta$ &  \hfill 0.120\ \ \   &  1.232  \\
 2 &  \hfill  0.074\ \ \   &   1.503  \\
 3 &   \hfill  0.094\ \ \   &   $ 1.677(1+0.0102Q^2 - 0.00084Q^4)$   \\ \hline
\end{tabular}
\vskip 0.5in

\begin{tabular}{|c c c c c c c|}
\hline
\multicolumn{7}{|p{6.3in}|}{Table III. Results \cite{linda} of a global fit  to the proton inelastic cross sections in units of $\mu$b and normalized to $[G_D(Q^2)]^2$.
These coefficients are defined in Eqs. 3, 5, and 8. As described in the text, each coefficient has a polynomial dependence on $Q^2$ which is in (GeV/c)$^2$. The global fit gives a $\chi^2$ per 523 degrees of freedom of 1.56.} \\  \hline
 & & & & & &\\
 $Q^{(2*n)}$ &  $\vert F_\Delta \vert_n^2$  & $A_{2n}$ &  $A_{3n}$ & $C_{1n}$ & $C_{2n}$ & $C_{3n}$ \\
 & & & & & &\\ \hline
 & & & & & &\\ 
$Q^0$ &\ \phantom{$-$}1.396E+01\ &\ $-$3.153E+00\ &\ $-$4.540E$-$01\ &\ \phantom{$-$}3.167E+02\ &\ $-$1.711E+01\ &\ $-$1.836E+02\  \\
$Q^2$ &\ $-$2.879E+00\ &\ \phantom{$-$}2.933E+01\ &\ \phantom{$-$}1.935E+01\ &\ \phantom{$-$}1.490E+03\ &\ $-$5.320E+03\ &\ \phantom{$-$}4.144E+03\  \\
$Q^4$ & \ \phantom{$-$}1.587E$-$01\  & & & \ \phantom{$-$}6.120E+02\ & \ \phantom{$-$}5.364E+03\ & \ $-$4.890E+03\  \\
$Q^6$ & & & & \   \phantom{$-$}1.823E+01\ & \  $-$1.224E+03\ & \   \phantom{$-$}1.229E+03\  \\
$Q^8$ & & & & \  $-$6.437E+00\ & \   \phantom{$-$}7.816E+01\ & \  $-$7.934E+01\  \\ \hline
\end{tabular}
\eject

\begin{tabular}{|c|c|c|}
\hline
\multicolumn{3}{|p{3.7in}|}{Table IV. NE11 and E133 $\Delta$(1232) transition form factor results determined from cross section data and normalized to $\mu_p G_D(Q^2)$. The first uncertainty is statistical, and the second is systematic including modeling and normalization uncertainties.} \\  \hline
 & & \\
 \ $Q^2$ (GeV/c)$^2$\ \  &   \ \ $F_\Delta(Q^2)/ \mu_p G_D(Q^2)$\  &\ Experiment\   \\
 & & \\ \hline
 & & \\ 
 1.64 &  \ \ 1.139$\pm$  0.007 $\pm$ 0.023\ \  &  NE11  \\ 
 2.34 &  \ \ 1.014$\pm$  0.007 $\pm$ 0.021\ \  &  NE11  \\ 
 3.05 &  \ \ 0.911$\pm$  0.008 $\pm$ 0.022\ \  &  NE11  \\ 
 3.75 &  \ \ 0.804$\pm$  0.008 $\pm$ 0.024\ \  &  NE11  \\ 
 3.86 &  \ \ 0.856$\pm$  0.014 $\pm$ 0.038\ \  &  NE11  \\ 
 4.82 &  \ \ 0.732$\pm$  0.020 $\pm$ 0.040\ \  &  NE11  \\ 
 5.79 &  \ \ 0.656$\pm$  0.024 $\pm$ 0.047\ \  &  NE11  \\ 
 6.75 &  \ \ 0.513$\pm$  0.052 $\pm$ 0.070\ \  &  NE11  \\ 
 2.41 &  \ \ 1.017$\pm$  0.009 $\pm$ 0.068\ \  &  E133  \\ 
 3.90 &  \ \ 0.773$\pm$  0.017 $\pm$ 0.069\ \  &  E133  \\ 
 5.87 &  \ \ 0.557$\pm$  0.017 $\pm$ 0.131\ \  &  E133  \\ 
 7.86 &  \ \ 0.431$\pm$  0.034 $\pm$ 0.174\ \  &  E133  \\ 
 9.83 &  \ \ 0.317$\pm$  0.046 $\pm$ 0.245\ \  &  E133  \\ \hline
\end{tabular}
\vskip 0.5in


\begin{tabular}{|c|c|c|c|c|c |}
\hline
\multicolumn{6}{|p{6.1in}|}{Table V. Ratios $\sigma_n/\sigma_p$ extracted from inelastic e-d cross sections in the region of the $\Delta$(1232) resonance for both resonant and non-resonant cross section components. The first uncertainty is statistical and the second is the quadratically combined systematic and modeling (dominant) uncertainties.  } \\  \hline
 & & & & &\\
 $Q^2$&  ${(\sigma_n/\sigma_p)}_{\Delta}$ & ${(\sigma_n/\sigma_p)}_{nres}$ & ${(\sigma_n/\sigma_p)}_{nres}$  & ${(\sigma_n/\sigma_p)}_{total}$ & Exp.\\
 (${\rm GeV\over c}$)$^2$ & Fit I & Fit I & Fit II& Fit III& \\
 & & & & & \\ \hline
 & & & & & \\ 
 1.64 &\ 0.81$\pm$0.10$\pm$0.05\ \ &\ 0.69$\pm$0.15$\pm$0.12\ \ &\ 0.41$\pm$0.04$\pm$0.15\ \ &\ 0.76$\pm$0.02$\pm$0.07\ \ &\ NE11\ \ \\ 
 2.34 &\ 0.73$\pm$0.13$\pm$0.09\ \ &\ 0.77$\pm$0.12$\pm$0.11\ \ &\ 0.53$\pm$0.04$\pm$0.14\ \ &\ 0.75$\pm$0.02$\pm$0.08\ \ &\ NE11\ \ \\ 
 3.05 &\ 0.63$\pm$0.22$\pm$0.10\ \ &\ 0.63$\pm$0.13$\pm$0.11\ \ &\ 0.43$\pm$0.04$\pm$0.13\ \ &\ 0.63$\pm$0.03$\pm$0.09\ \ &\ NE11\ \ \\ 
 3.75 &\ 0.14$\pm$0.35$\pm$0.21\ \ &\ 0.69$\pm$0.14$\pm$0.10\ \ &\ 0.37$\pm$0.04$\pm$0.12\ \ &\ 0.54$\pm$0.03$\pm$0.10\ \ &\ NE11\ \ \\ 
 2.41 &\ 0.71$\pm$0.05$\pm$0.16\ \ &\ 0.53$\pm$0.02$\pm$0.17\ \ &\ 0.45$\pm$0.02$\pm$0.15\ \ &\ 0.57$\pm$0.01$\pm$0.11\ \ &\ E133\ \ \\ 
 3.90 &\ 0.61$\pm$0.10$\pm$0.21\ \ &\ 0.45$\pm$0.03$\pm$0.13\ \ &\ 0.36$\pm$0.02$\pm$0.12\ \ &\ 0.48$\pm$0.01$\pm$0.10\ \ &\ E133\ \ \\ 
 5.87 &\ 0.83$\pm$0.24$\pm$0.43\ \ &\ 0.43$\pm$0.02$\pm$0.13\ \ &\ 0.42$\pm$0.02$\pm$0.11\ \ &\ 0.46$\pm$0.01$\pm$0.11\ \ &\ E133\ \ \\ 
 7.86 &\ 1.03$\pm$1.02$\pm$0.92\ \ &\ 0.46$\pm$0.04$\pm$0.13\ \ &\ 0.47$\pm$0.02$\pm$0.11\ \ &\ 0.48$\pm$0.02$\pm$0.11\ \ &\ E133\ \ \\ \hline
\end{tabular}
\eject

\baselineskip 14pt

\includegraphics{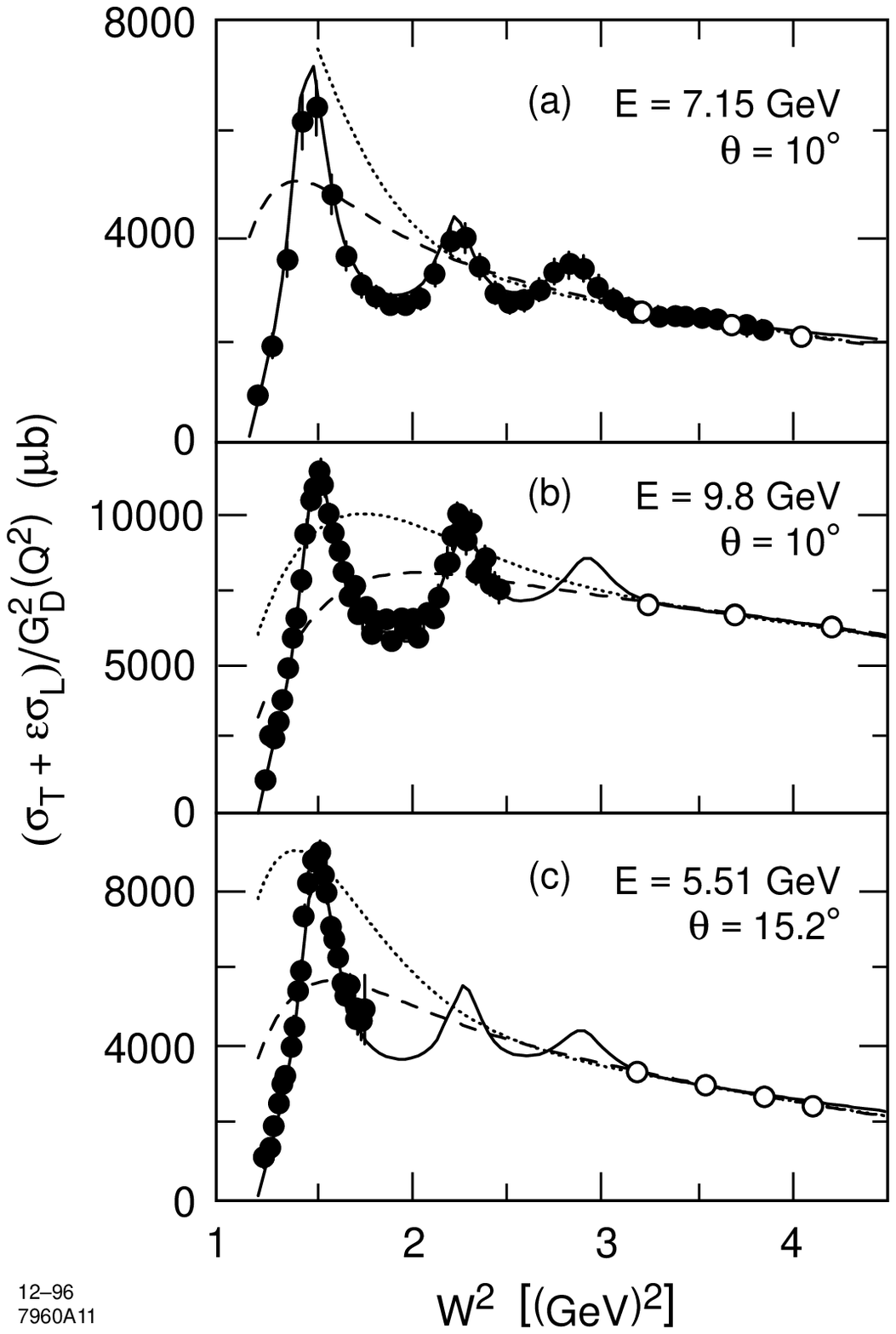}
\vbox to 7.3in{}
\noindent
Fig. 1.  Three sample spectra displaying both resonance and deep inelastic photoabsorption cross section data. The data are normalized to $G_D^2(Q^2)$. Resonance data ($\bullet$) are shown from\cite{brasse} (a), \cite{rock} (b), and this experiment (c). The deep inelastic data ($\circ$) from \cite{whitlowa} are from nearby kinematics and were bin-centered to the indicated kinematics using the global fit which is shown as a solid line. Also shown are deep inelastic global fits given by the dashed curve \cite{whitlowb} and the dotted curve \cite{amaudruz}. Both of these used the same parameterization for $R$ \cite{whitlowc}.
\eject

\includegraphics{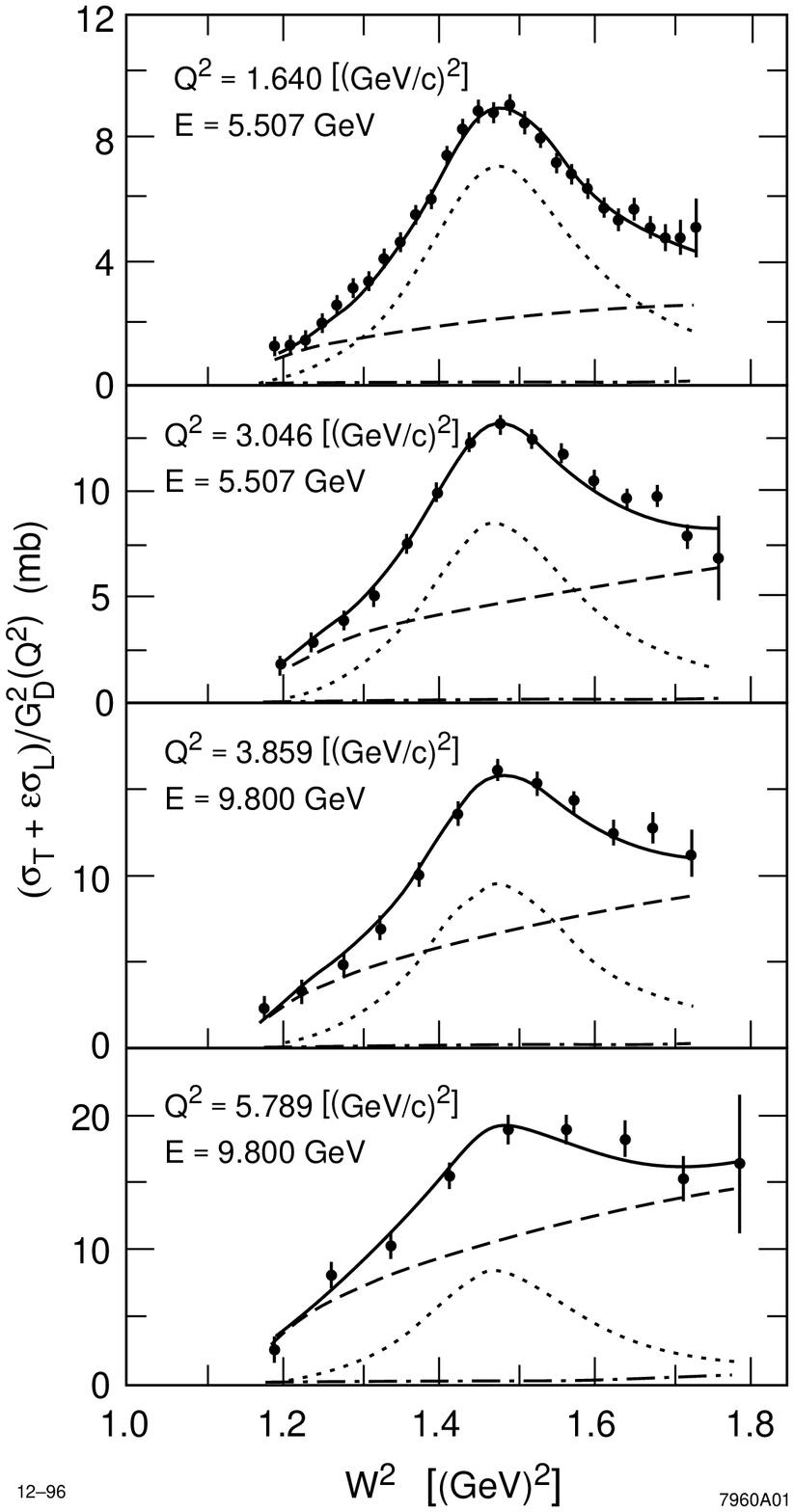}
\vbox to 7.3in{}
\noindent
Fig. 2.  Sample virtual photoabsorption cross-section spectra, for the proton measured in this experiment. The errors include statistical and point-to-point systematic uncertainties. The contributions to the spectra from the higher-mass resonance (dot-dash) and the nonresonant (dashed) background were determined from the global fit to the data. The $\Delta(1232)$ strength (dotted) was determined using a single parameter fit to determine $F_\Delta(Q^2)$ for each spectrum. Also shown is the sum (solid) of these cross section components. The indicated $Q^2$ is at the $\Delta (1232)$ resonance mass of 1.232 GeV.

\eject

\includegraphics{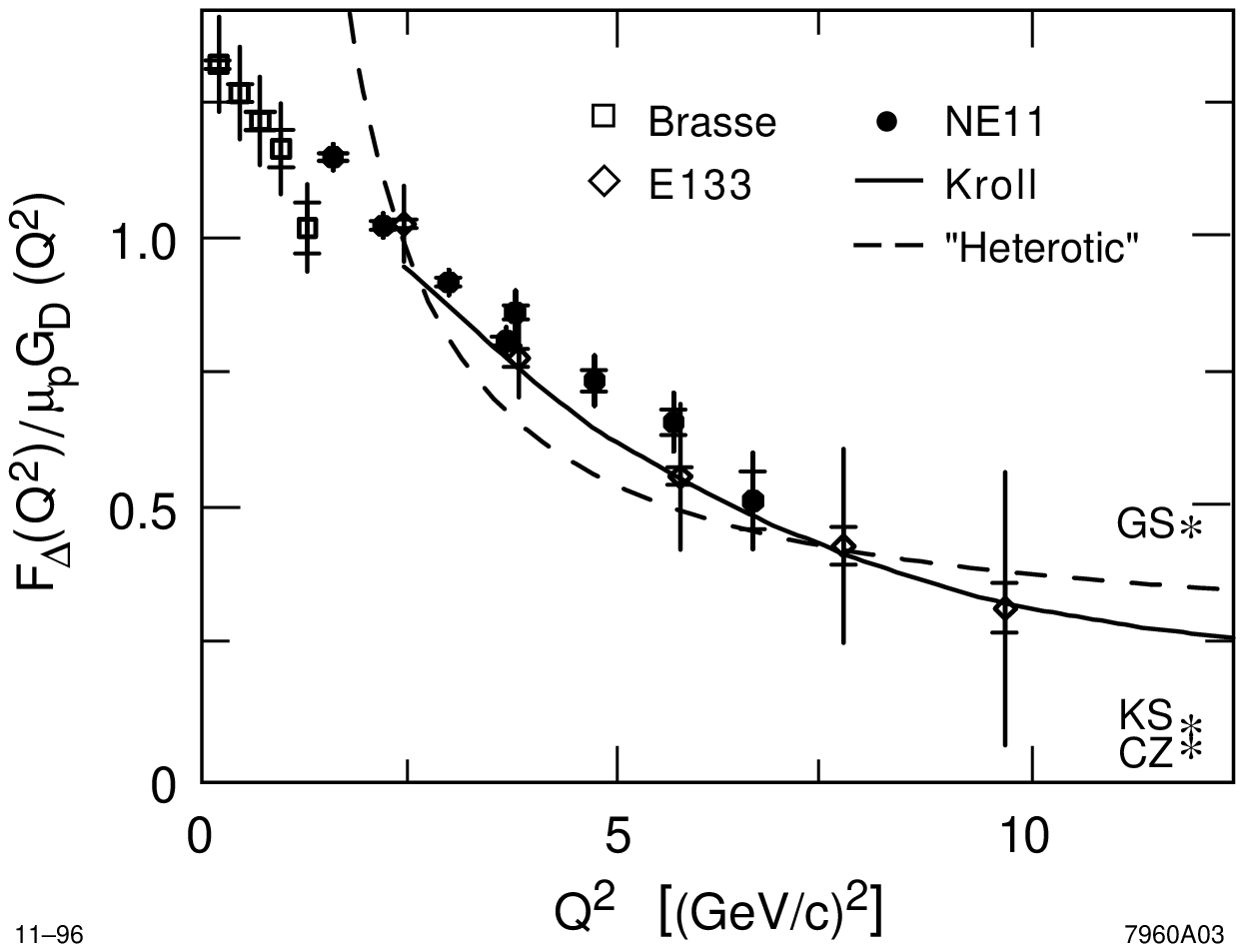}
\vbox to 5.0in{}
\noindent
Fig. 3.  Extracted $\Delta$(1232) transition form factors from fits to individual $ep$ cross-section spectra at each $Q^2$ point using the global fit to describe the nonresonant background. The inner error bars are statistical, and the outer error bars are total errors including systematic, modeling and normalization uncertainties. The diquark model fit due to Kroll, {\it et al.}\cite{krolla} is shown as well as the heterotic prediction from Stefanis and Bergmann \cite{stefberg} and the three asymptotic predictions (denoted by $*$ and labeled by GS, KS, and CZ) due to Carlson and Poor \cite{carlsonb}, which have been evaluated at $Q^2 = 12$ (GeV/c)$^2$.                        
\eject

\includegraphics{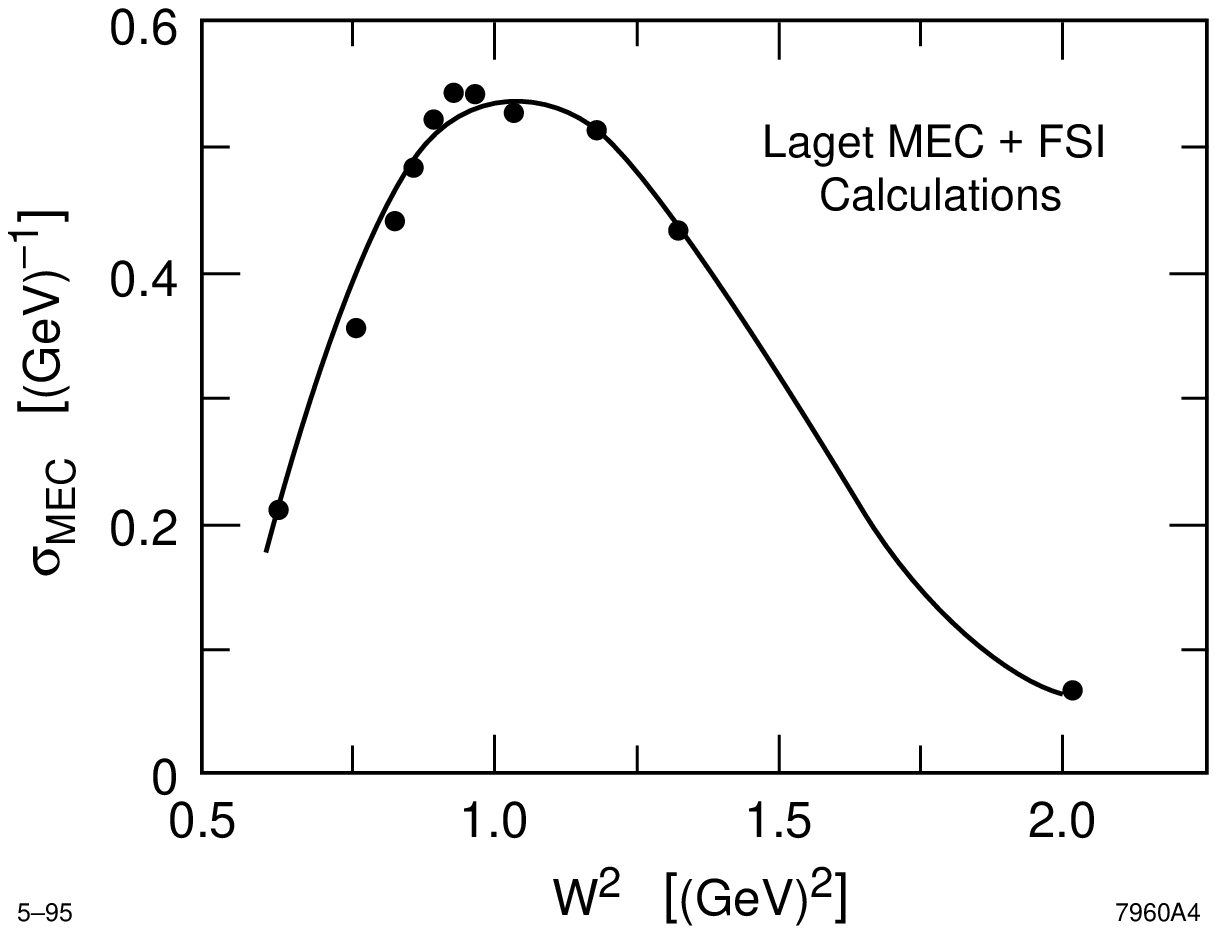}
\vbox to 5.2in{}
\noindent
Fig. 4.  Reduced cross section (see Eq. 9) calculations by Laget \cite{laget} (points) at $Q^2$ = 1.75 (GeV/c)$^2$ of MEC and FSI contributions to the deuteron inelastic cross section in the $\Delta(1232)$ region. These points were determined from the difference of two cross section calculations. The curve is a simple polynomial fit.      
\eject

\includegraphics{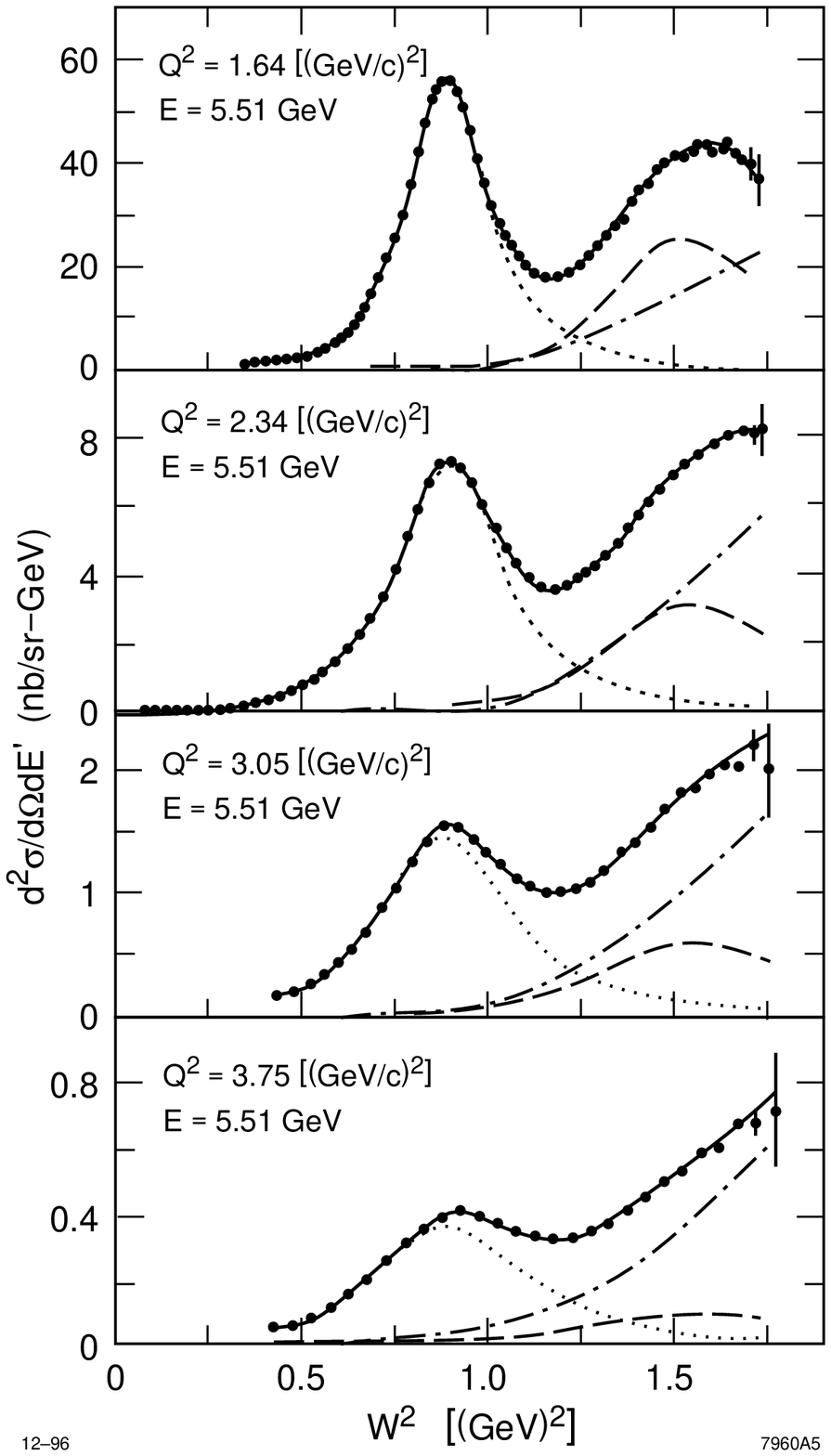}
\vbox to 7.0in{}
\noindent
Fig. 5.  Inelastic $ed$ cross sections per nucleon from this experiment measured in the $\Delta$(1232) resonance region for the kinematic points indicated. The indicated $Q^2$ are at the $\Delta (1232)$ resonance mass of 1.232 GeV. The errors include statistical and point-to-point systematic uncertainties. Also shown are the quasielastic (dotted), the $\Delta$(1232) resonance (dashed), and the inelastic nonresonant (dot-dash) contributions obtained from Fit I. Contributions from the higher mass resonance region and MEC were included in the fit, but are not shown because they are too small to be seen clearly on this scale. The sum of all these contributions is shown as the solid curve.
\eject

\includegraphics{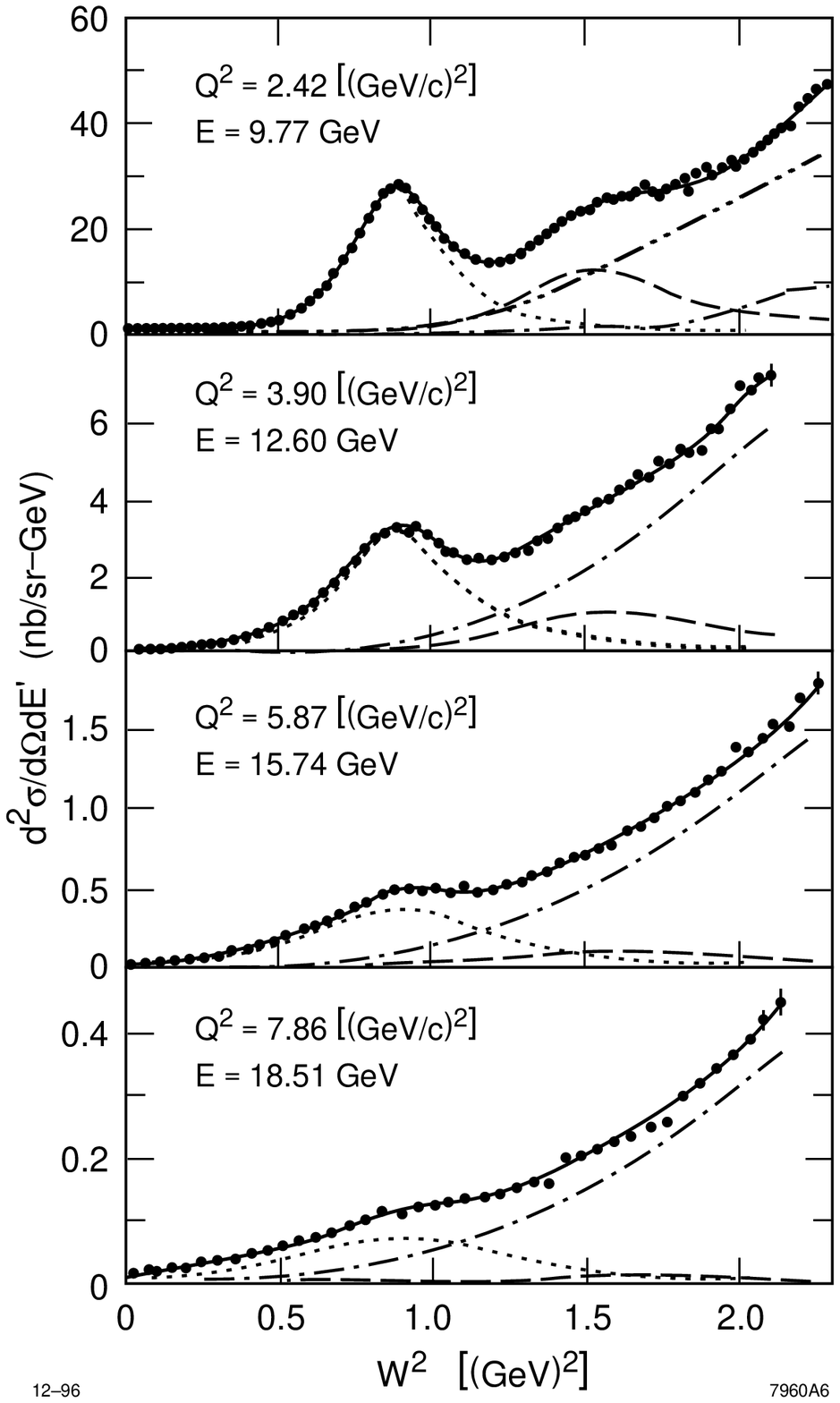}
\vbox to 7.3in{}
\noindent
Fig. 6.  Inelastic $ed$ cross sections from experiment E133 \cite{rock} measured in the $\Delta$(1232) resonance region for the kinematic points indicated. The curves have the same meaning as those in Fig. 5. The indicated $Q^2$ are at the $\Delta (1232)$ resonance mass of 1.232 GeV.
\eject

\includegraphics{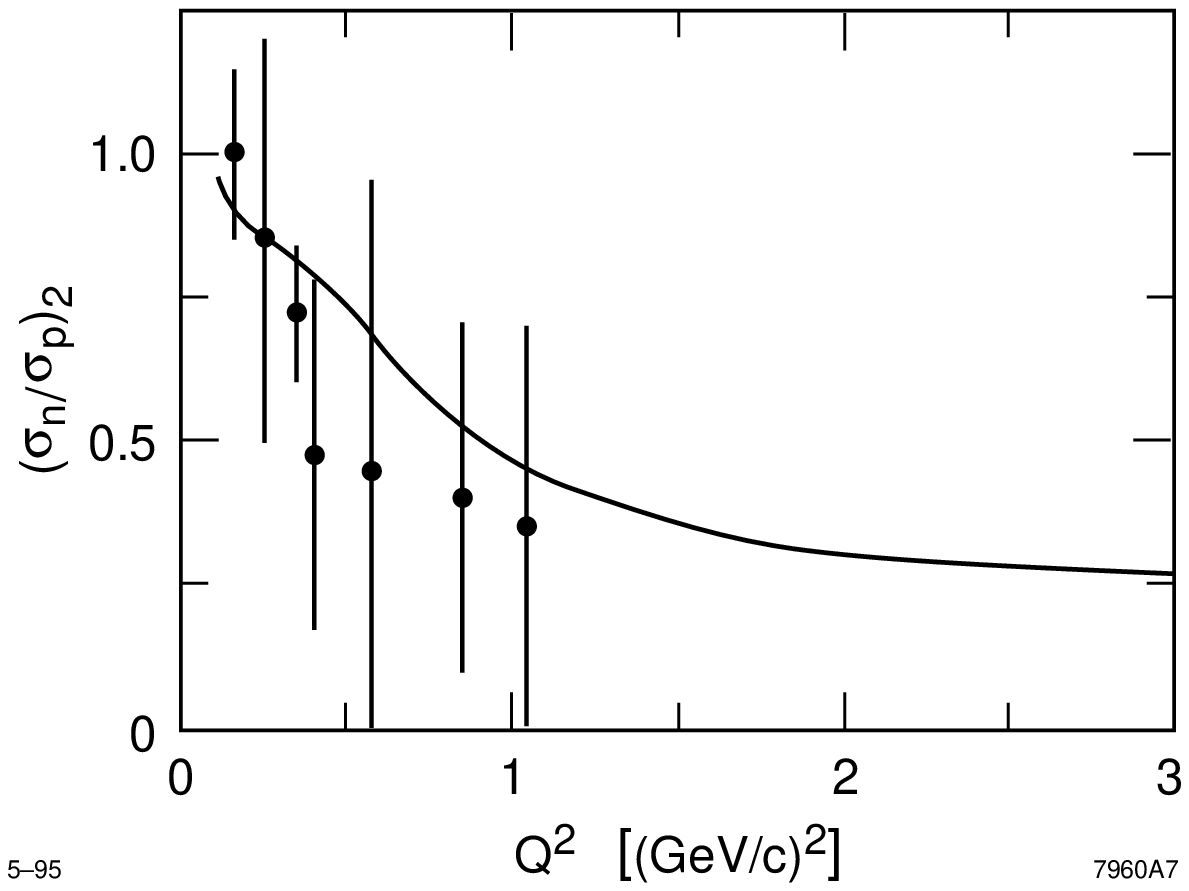}
\vbox to 5.1in{}
\noindent
Fig. 7.  Previous results \cite{kobb} at low Q$^2$ for the higher-mass resonance ratio, $(\sigma_n/\sigma_p)_{2}$. The curve is a relativistic constituent quark model prediction \cite{warns} for the S$_{11}(1535)$ resonance.
\eject                                                                        

\includegraphics{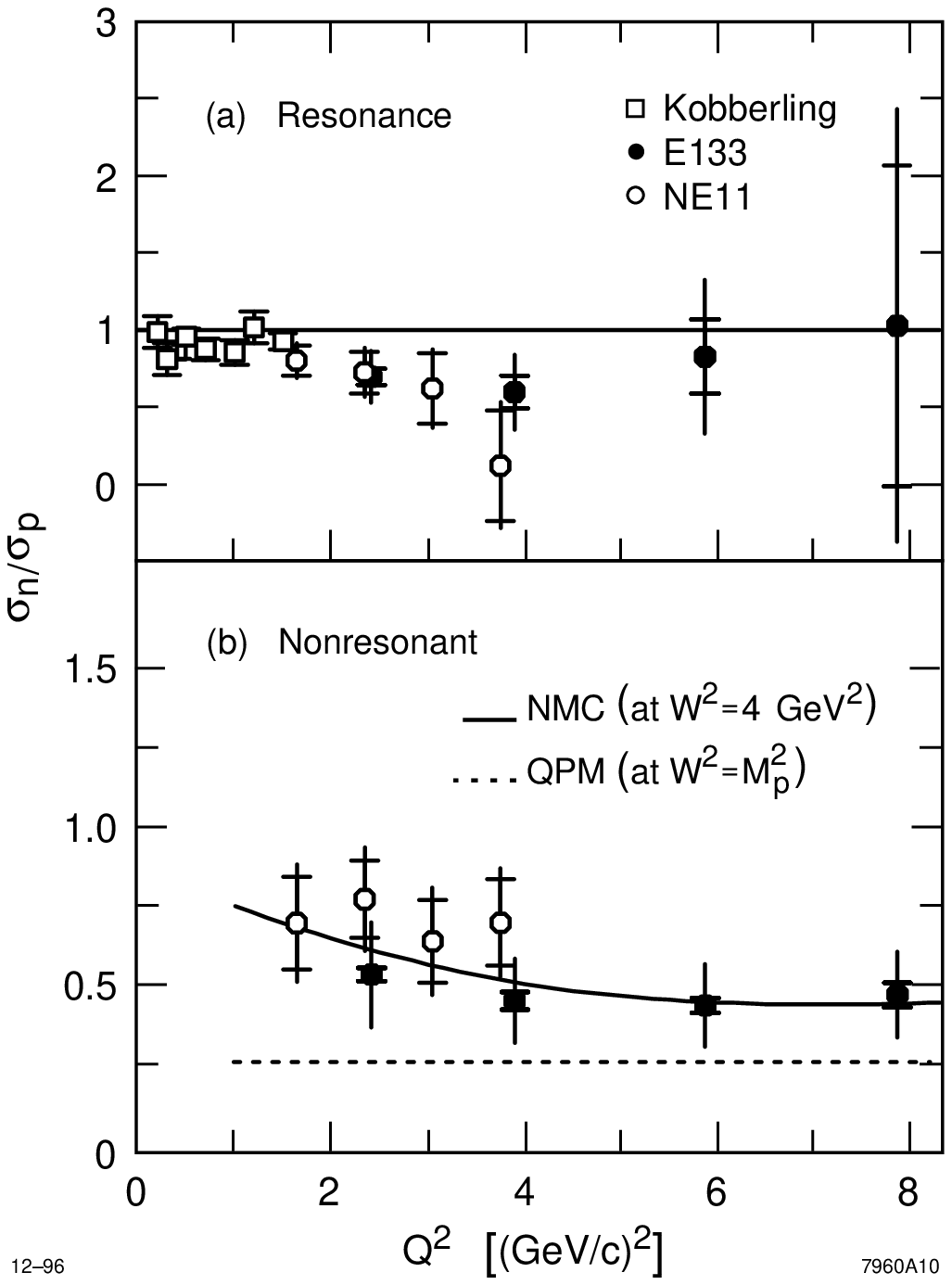}
\vbox to 7.1in{}
\noindent
Fig. 8.  The ratios (a) ${(\sigma_n/\sigma_p)}_{\Delta}$ and (b) ${(\sigma_n/\sigma_p)}_{nres}$ from Fit I. The inner error bar is statistical and the outer error bar is systematic including modeling uncertainties. These results were found using the ``standard'' model choices: Paris deuteron wave function, smearing model SM-I, off-shell correction OS-IV, and $(\sigma_n/\sigma_p)_{2}$ = 0.3. Previous data at low Q$^2$ from K\"obberling {\it et al.} \cite{kobb} are also shown. The solid curve in (b) was determined from \cite{amaudruz,whitlowc}.

\eject                                                                        

\includegraphics{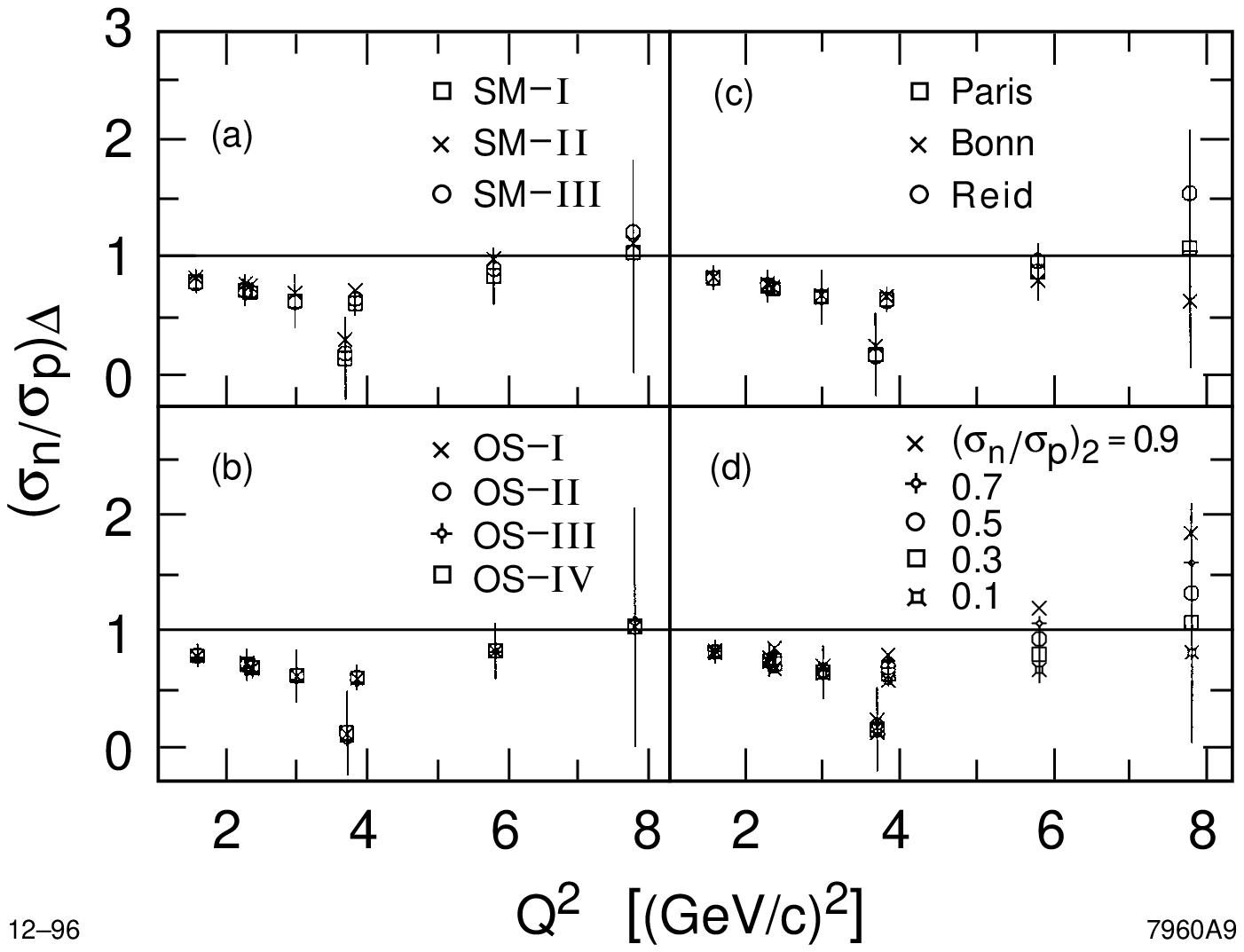}
\vbox to 5.0in{}
\noindent
Fig. 9.  The ratio $\sigma_n/\sigma_p$ for $\Delta$(1232) resonance cross sections for several (a) smearing models, (b) off-shell corrections, (c) deuteron wave functions, and (d) choices of the parameter $(\sigma_n/\sigma_p)_{2}$. Only statistical error bars for the ``standard'' results (defined in text and in Fig. 8 and denoted by squares) are shown.
\eject

\includegraphics{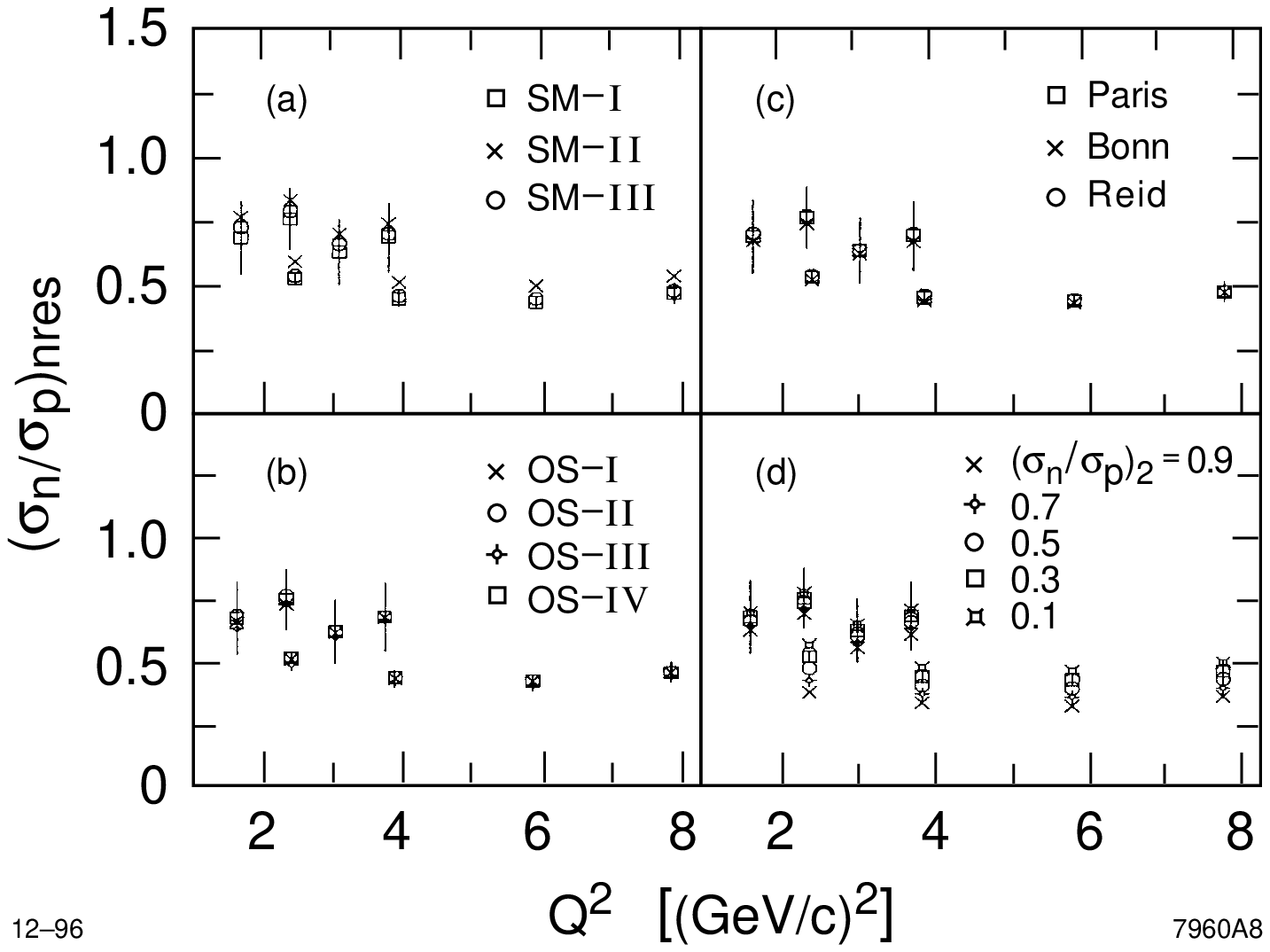}
\vbox to 5.1in{}
\noindent
Fig. 10.  The ratio $\sigma_n/\sigma_p$ for the inelastic nonresonant cross sections in the $\Delta$(1232) resonance region for several (a) smearing models, (b) off-shell corrections, (c) deuteron wave functions, and (d) choices of the parameter $(\sigma_n/\sigma_p)_{2}$. Only statistical error bars for the ``standard'' results (defined in text and denoted by squares) are shown.
\eject

\end{document}